\documentclass[conference]{IEEEtran}

\usepackage[american]{babel}
\usepackage{cite}
\usepackage[hyphens]{url}
\usepackage{color}
\usepackage{booktabs}
\usepackage{rotating}
\usepackage{subfigure}
\usepackage{multirow,tabularx}
\usepackage{flushend}
\usepackage{pifont}
\usepackage{amsmath, amssymb}
\usepackage{hyphenat}
\usepackage[table]{xcolor}

\newcommand{\etAl}{et\,al.}
\newcommand{\eg}{e.\,g., }
\newcommand{\ie}{i.\,e., }

\begin{document}

\rowcolors{1}{gray!25}{white} 
 
% paper title
\title{A Study of Newly Observed Hostnames \\ and DNS Tunneling in the Wild}

\author{
\IEEEauthorblockN{Dennis Tatang,
Florian Quinkert,
Nico Dolecki, 
and Thorsten Holz}
\IEEEauthorblockA{
\textit {\{firstname.lastname\}@rub.de} \\
Ruhr University Bochum, Germany}
}

\maketitle

\begin{abstract}
The domain name system (DNS) is a crucial backbone of the Internet and millions of new domains are created on a daily basis. While the vast majority
of these domains are legitimate, adversaries also register new hostnames to carry out nefarious purposes, such as scams, phishing, or other types of attacks. 
In this paper, we present insights on the global utilization of DNS through a measurement study examining exclusively \emph{newly observed hostnames} via passive DNS data analysis. We analyzed more than two billion such hostnames collected over a period of two months. Surprisingly, we find that only three second-level domains are responsible for more than half of all newly observed hostnames every day. More specifically, we found that Google's \emph{Accelerated Mobile Pages} (AMP) project, the music streaming service Spotify, and a \emph{DNS tunnel} provider generate the majority of new domains on the Internet. DNS tunneling is a covert channel technique to transfer arbitrary information over DNS via DNS queries and answers. This technique is often (ab)used by attackers to transfer data in a stealthy way, bypassing traditional network security systems. We find that potential DNS tunnels cause a significant fraction of the global DNS requests for new hostnames: our analysis reveals that nearly all resource record type NULL requests and more than a third of all TXT requests can be attributed to DNS tunnels.

Motivated by these empirical measurement results, we propose and implement a method to identify DNS tunnels via a step-wise filtering approach that relies on general characteristics of such tunnels (e.g.,  number of subdomains or resource record type).
Using our approach on empirical data, we successfully identified 273 suspicious domains related to DNS tunnels, including two known APT campaigns (\emph{Wekby} and \emph{APT32}).
\end{abstract}


\section{Introduction}
\label{sec:intro}

The resolution of domain names to IP addresses provided by the Domain Name System (DNS) is fundamental for comfortably using the Internet.
Every Internet user utilizes this functionality, thus making it an attractive target for attacks.
As a result, it is important to understand the development and use of DNS in the wild.
Abuses such as DNS as amplification protocol in the context of DDoS attacks or cache poisoning attacks are known and have been thoroughly analyzed in previous publications~\cite{kambourakis2007detecting,son2010hitchhiker}.
In addition, various measurement studies described the development and changes in the DNS ecosystem and discussed several aspects, such as interception, censorship, dependencies, or measurement challenges~\cite{217551,pearce2017global,Dell'Amico:2017:LMM:3134600.3134637,7460220}.
However, a comprehensive analysis of previously unknown or new requested hostnames has not been performed so far. 

In this paper, we conduct a systematic measurement study on this topic on passive DNS data obtained from the globally distributed Farsight DNS sensor network~\cite{farsight_noh}.
Our analyzed data set consists of newly observed fully qualified domain names (FQDNs) only, i.e., it does not contain widely known domain names like \textit{google.com} or \textit{facebook.com}, but only domains observed being resolved for the very first time. In total, we analyzed more than two billion such domains collected over a period of two months.

We found that the majority of these FQDNs do not originate from an average user surfing the Internet, but are automatically generated.
In a first step, we performed an in-depth structural analysis of the obtained FQDNs to understand which application scenarios require the use of new FQDNs and later on analyze them in detail.
We found especially automated requests from Google's AMP project, Spotify, and DNS tunnels in our data set responsible for half of all entries, indicating further analysis is crucial.
From a security perspective, especially DNS tunneling is interesting because it allows an attacker the covert transfer of information.
Although, many publications already dealt with DNS tunnels~\cite{paxson,qi2013bigram,homem2017harnessing, aiello2015dns,farnham2013detecting,dusi2009tunnel,sheridan2015detection,DBLP:journals/corr/abs-1004-4358,nuojua2017dns,satam2015anomaly,born2010ngviz,ellens2013flow,cejka2014stream,karasaridis2006,aiello2013basic}, a comprehensive global overview of the real-world usage of DNS tunnels is missing. 
Therefore, we analyze to what extent DNS tunnels can be found in a large, aggregated data set of newly observed hostnames. Furthermore, we search for examples of malicious activity and conclude that it is an actual real-world threat. 

As previously explained, DNS tunnels are hidden, often not monitored communication channels.
Attackers use them for the extraction of information as well as the establishment of command and control channels (e.g., FrameworkPOS~\cite{frameworkpos} or C3PRO-RACCOON~\cite{C3PRO-RACCOON}).
Even advanced persistent threat (APT) actors use this technique to successfully attack their targets (e.g., Wekby~\cite{wekby}, APT32~\cite{apt32}, or APT34~\cite{apt34}).
The most recent example of using a DNS tunnel by malware is from February 2018, a point of sale (POS) malware used it for data exfiltration (UDPoS~\cite{udpos, udpos_first}).
Although the technique is already known for some time, it is still popular as an attack vector~\cite{pcr123} and therefore it is important to understand usage in order to identify campaigns that use this technique early on.
Existing efforts to analyze DNS tunnels depend on an internal network view, \ie a local network in which the presence of DNS tunnels is detected and analyzed.
However, an overview on globally distributed sensor data is not possible with these systems as these approaches use single attributes that would generate high false positives on such aggregated passive DNS data, e.g., due to CDNs. 
In contrast, our approach of examining passive DNS data with newly observed hostnames from a distributed sensor network allows a broad overview of DNS tunnel usage.
In particular, we introduce a step-wise approach with filter functions which take characteristics of known DNS tunnels into account to reduce the passive DNS data down to potential DNS tunnel domains, e.g., number of subdomains per second-level domain, used resource record type (e.g., A, TXT, NULL), or level of full hostnames.
Thus, we can analyze the filtered data to understand the extent such tunnels are used in the wild.

We again analyzed more than two billion passive DNS entries and discovered 273 candidate domains within resource record types NULL and TXT, which were potentially used for DNS tunneling. We observed that almost all type NULL traffic and about 35 percent of type TXT traffic is related to DNS tunnels.
Additionally, we provide a survey of the development of DNS tunnel usage by malicious software.
With our analysis approach, we were able to identify two APT groups (APT32 and Wekby) related to ten second-level domains in our data set, which we analyze in more detail in two separate case studies.
The Wekby case study proves the importance of monitoring even old DNS tunnel domains. 
In that specific case, we identified a DNS tunnel belonging to an APT campaign that has been featured in blog posts back in the year 2016. 
Nevertheless, we detected activity in our gathered data, which means that the old infrastructure was still in use at a much later date. 
Finally, we discuss threats to validity of our filtering approach.

\smallskip \noindent
In summary, we make the following contributions:
\begin{enumerate}
\item We conduct a measurement study of the usage of DNS requests with new fully qualified domain names on a passive DNS data set.
\item We provide insights on how DNS tunnels are used in practice and propose a simple, yet effective collection of filtering functions for identifying DNS tunnels in passive DNS data (or rather identifying suspicious domains) and demonstrate its applicability in practice.
\item We discuss two case studies of APT campaigns using DNS tunnels (APT32 and Wekby) seen in our collected data set and present a brief survey of malware utilizing DNS tunneling techniques.
\end{enumerate}

In the remainder of this paper, we first introduce technical background information in Section~\ref{sec:background}. 
Afterwards, we present our measurement study in Section~\ref{sec:measurement}, followed by introducing our approach to identify suspicious domain names in Section~\ref{sec:approach}.
In Section~\ref{sec:evaluation}, we present the usage of our filter functions and demonstrate that the resulting candidate domains are indeed domains used by DNS tunnels.
In this data set, we also discovered domains associated with APT campaigns and we analyze these findings in two case studies in Section~\ref{sec:casestudies}.
Subsequent, we discuss limitations of our work in Section~\ref{sec:limitations}, review related work in Section~\ref{sec:related_work}, discuss future work in Section~\ref{sec:futurework}, and finally conclude in Section~\ref{sec:conclusion}.
\section{Background}
\label{sec:background}

Before we present our measurement study, we provide basic information to ease understanding the rest of our paper. 
First, we describe the DNS and passive DNS data.
Afterwards, we introduce the concept of DNS tunnels.

\subsection{Domain Name System}

\begin{figure}
\centering
	\includegraphics[width=0.8\linewidth]{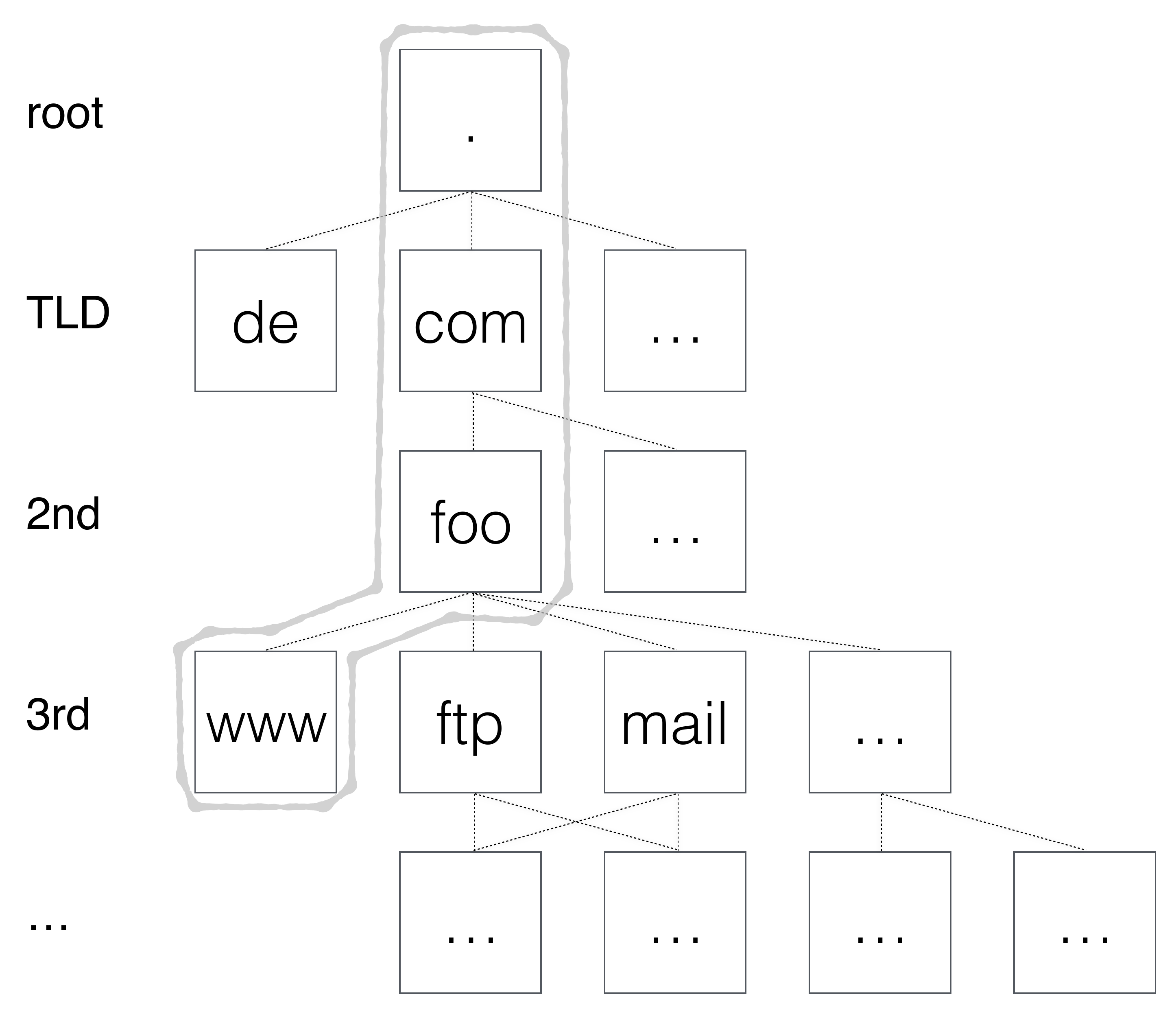}
	\caption{Domain name hierarchy root tree}
	\label{fig:domain_name_hierarchy}
\end{figure}

The Domain Name System (DNS) is hierarchically structured so that no central database with all DNS information exists.
When a client needs information from the DNS, it sends a request to a predefined local DNS server.
If this server cannot answer the request, it forwards the request to one of the root servers. Then the request is forwarded to the server of the top-level domain, which forwards the request to the server responsible for the second-level domain. This continues until a DNS server can provide the appropriate answer.
Servers forwarding a DNS request are referred to as \emph{recursive DNS servers}.
Accordingly, it is possible to visualize the DNS namespace as a tree~\cite{rfc1035} (see Figure~\ref{fig:domain_name_hierarchy}).
The most right part of a domain is at the topmost position in the hierarchy of the tree (.\textit{[empty]}) and the most left part is at the lowest position (\eg www).
The highest level is called the root.
Topologically below, and thus listed to the left of the root, is the name of a top-level domain (TLD) (\eg com).
Below the top-level domain are the names of the second-level domains (\eg foo) followed by third-level domains or simply further labels of lower levels.
Each level in a domain is called \emph{label}.
The full name of a domain is called Fully Qualified Domain Name (FQDN).
Hence, the domain name \textit{www.foo.com.} is an FQDN with three levels.
A subdomain is a part of an FQDN, e.g., \textit{example.www.foo.com} is a subdomain of \textit{www.foo.com}. 
An FQDN's maximum length is restricted to 256 characters, effectively it is still necessary to remove the TLD (at least two characters) and the root (1 character), allowing a maximum number of 253 characters. 
The maximum length of individual labels is defined by 63 characters.

Besides translating memorable domain names into their corresponding IP addresses, DNS offers further features.
Each DNS request contains an information called \emph{resource record type} (rrtype) which encodes the purpose of the corresponding DNS request:
type A and type AAAA resolve domains to IPv4 or IPv6 addresses, respectively, while type CNAME provides aliases. Type MX is used to find the matching mail server and type NS returns the corresponding nameserver.
Other types include, \eg TXT for transmitting text data and NULL for arbitrary content. 
In total, DNS supports 92 different resource record types~\cite{dns_parameters}.

\subsection{Passive DNS}

Passive DNS (pDNS) was commercialized in 2002 by Sandstorm Enterprises in the NetIntercept product which appears in the work of Corey \etAl~\cite{corey}.
In 2004, Weimer introduced the concept of pDNS as a defense against malware~\cite{weimer2005passive}. This concept works as follows:
Recursive DNS servers log requests they receive from other DNS servers.
Passive DNS replicates the received requests from multiple recursive DNS servers into a central database.
In other words, the overall result is aggregated data.
Later on, researchers and analysts can use pDNS databases, \eg to discover DNS queries resolved for a particular domain name, corresponding nameservers or other zones using the same nameservers.
This provides an opportunity to search for known malicious IP addresses and find all domain names associated with these IP addresses.

\begin{figure}
\centering
  \includegraphics[width=\linewidth]{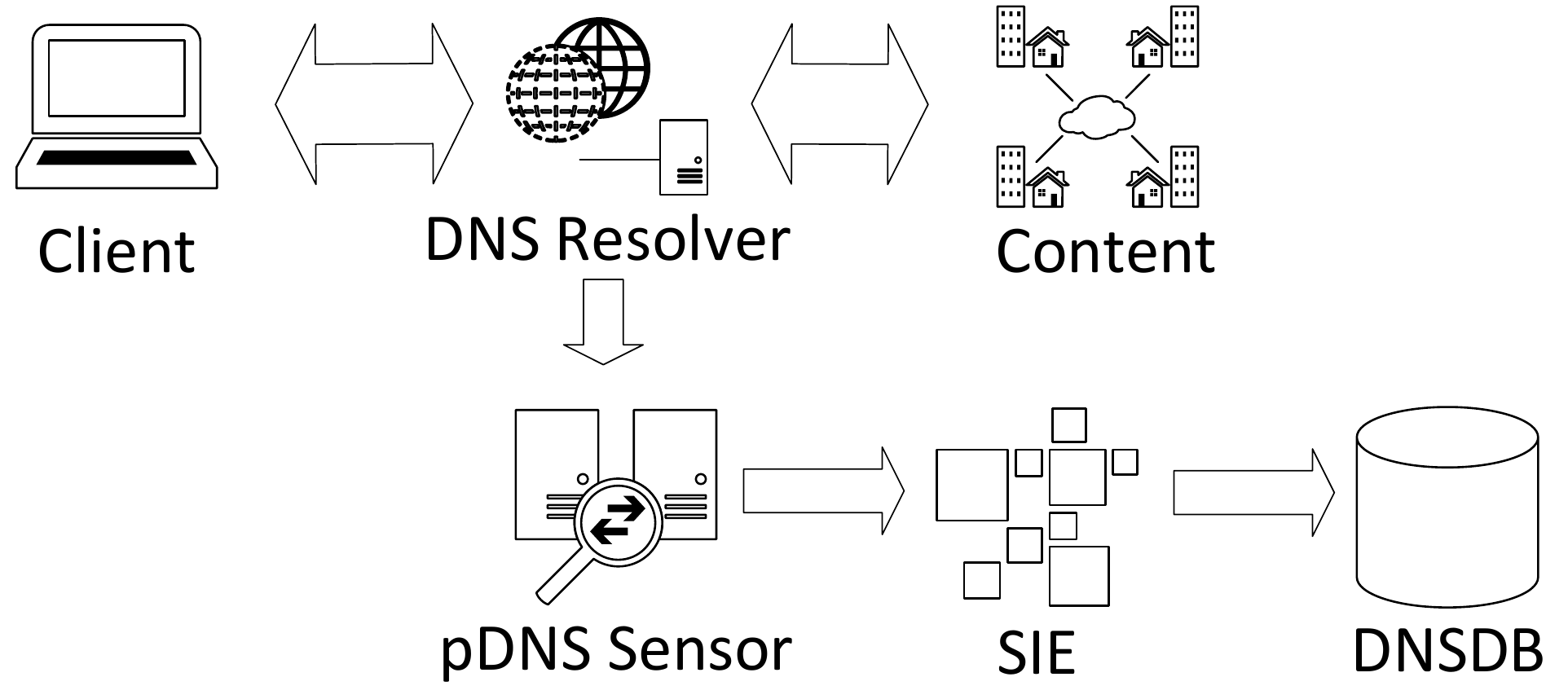}
  \caption{Structure of Farsight Passive DNS}
  \label{fig:pdns_structure}
\end{figure}

Various companies collect data from recursive DNS servers (in this context often referred to as pDNS sensors) in large databases. 
For example, Farsight operates a globally distributed passive DNS sensor network, collects the data centrally (DNSDB), and provides access to it via live feeds (Security Information Exchange (SIE))~\cite{farsight_noh}.
The advantage of these live feeds is that the raw data can be saved, including all seen DNS requests, but also prefiltered data, \eg only new FQDNs that makes it comfortable for further analysis.
Figure~\ref{fig:pdns_structure} shows the structure of Farsight pDNS.
We expect Farsight to receive a significant fraction of all DNS requests observable in the wild due to the worldwide distribution of pDNS sensors~\cite{gao2013empirical}.

A pDNS entry contains various information like a timestamp and a message field.
Table~\ref{tab:pdns_message_example} presents an exemplary pDNS message field from the second-level domain \textit{teriava.com}. 
It is a dictionary including all relevant information, in particular: 

\begin{enumerate}
\item Domain field: the used second-level domain.
\item rrname field: the FQDN, i.e., the domain with all subdomains.
\item rrtype field: resource record type of the DNS request.
\item rdata field: the information for the DNS request response.
\end{enumerate}

The bailiwick field indicates the authoritative server~\cite{rfc7719}.
It is used by Farsight to avoid falsely accepting DNS results from untrustworthy sources.
The other fields are not necessary for the further course of our work.

\begin{table}[tbp]
	\centering
	\caption{pDNS message dictionary example (we omitted the fields keys and new\_rr since their values are empty in our example)}
	\label{tab:pdns_message_example}
	\begin{tabular}{@{}ll@{}}
		\toprule
		field      & value                   \\ \midrule
		domain     & teriava.com.            \\
		time\_seen & 2017-07-01 09:35:04     \\
		bailiwick  & teriava.com.            \\
		rrname     & dsu9jr2czl.teriava.com. \\
		rrclass    & IN                      \\ 
		rrtype     & A                       \\
		rdata      & {[}"127.0.0.1"{]}       \\ \bottomrule
%		keys       & {[}{]}                  \\
%		new\_rr    & {[}{]}                  \\ \bottomrule
	\end{tabular}
\end{table}

\subsection{DNS Tunneling} 
Besides the primary purpose of the DNS protocol, namely to query different types of data related to a specific domain, it is possible to use the hierarchical infrastructure to send data over it.
The DNS requests of the queried domains go through the recursive hierarchy of the DNS up to the authoritative nameserver.
A requirement to use DNS tunnels is the access to a domain and a DNS server (authoritative nameserver), which receives the DNS requests for the domain.
The admin of an authoritative nameserver can observe all incoming DNS queries.
Therefore, the answers to the queries are under control of that admin, too.
This behavior offers the admin a way to receive and send data (data exfiltration/infiltration), \ie to establish a two-way communication channel.
In particular, one-way communication (upstream) can be particularly hard to detect since it may be used very stealthy.
The advantages of DNS tunneling include that DNS is almost always available, no direct connection is established between victim and attacker, and pure data exfiltration (upstream only) is difficult to detect.

Note, we focus on DNS tunnels transferring data inside hostnames. Our research in Section~\ref{sec:approach} (as well as previously known malware see Section~\ref{subsec:known_malware} and the use of various DNS tunnel tools see Section~\ref{subsec:structure_analysis}) showed that this type of tunnel is most common in practice, and we thus concentrated on this technique in the rest of this paper.

DNS tunnels have two closely related main purposes.
First, establishing a communication channel between two hosts which are not allowed to communicate with each other.
Second, exchanging information in an obfuscated way.
Many public networks require their customers to login before surfing and use DNS to display a captive portal.
The availability of DNS enables a customer to use a DNS tunnel and establish a connection with a DNS server under her control to surf the Internet.
Even worse, an intruder can use a DNS tunnel in an internal network to exfiltrate information, such as passwords, or receive commands from an outside server.
Since DNS is often not monitored, this way of exchanging information often remains undetected and has already been successfully used by malware (see Sections~\ref{subsec:known_malware} and~\ref{sec:casestudies}).

\section{Measurement Study on the Usage of New Fully Qualified Domain Names (FQDNs)}
\label{sec:measurement}

In the following, we present the results of a measurement study of newly observed hostnames to understand the possibilities of pDNS data analysis.
Thereby, we focus on DNS requests with new fully qualified domain names (FQDN) only. 
We want to explore the reasons for requests with new hostnames since these are not conventional resolutions generated by a user surfing the Internet. Additionally, we analyze the  distributions of resource record types and the utilization of second-level domains with most subdomains.

\subsection{Data Set Description}

Farsight provided us access to their data live feed (channel 213).
This feed is pre-filtered in terms of it is processing only newly observed hostnames (FQDNs).
In other words it means we only see FQDNs that have not been observed by Farsight before. However, of course we see also already known second-level domains such like \textit{ampproject.net}. The term new refers to the full hostnames (e.g., \textit{new.example.ampproject.net}).
We stored the live feed for about two months between June 30th, 2017 and September 1st, 2017.
In July and August, more than two billion (2,041,665,066) pDNS entries were collected and saved ($\sim$800GBytes).
The mean count per day is 32,930,081.71, the median is 34,374,936.5, and the standard deviation is 3,171,327.81.
The data set is large despite the limitation to new FQDNs, but still tiny compared to all DNS requests in total on the Internet, \eg Google DNS servers receive about 400 billion requests a day~\cite{google_dns}.
Our later performed analyses are thus supported by our data set, which is prefiltered by Farsight.

\subsection{General Measurement Results}

Our analysis starts with statistics on various information that can be obtained via pDNS data and then we present an in-depth analysis with enriched network and manual search information.

\subsubsection*{Distribution resource record types (rrtype)}
Table~\ref{tab:distribution_rrtype} sums up the pDNS entry counts by rrtype and the share related to the total observed data between June and August 2017.
It is noteworthy that in total 21 different rrtypes occur, although the primary task of DNS is the resolution of hostnames into IP addresses, represented by rrtype A and AAAA, for IPv4 and IPv6, respectively. 
Type A entries make up just over half of all entries (almost 55~\%). 
The proportion of AAAA rrtype entries is rather low with almost 10~\%.
More than 95~\% of the entries represent five different types (A, NULL, AAAA, CNAME, TXT, ordered by frequency).
The remaining amount is distributed among the types NS, MX and others, including SOA, WKS, PTR, DNAME, RP, HINFO, SRV, SPF, NAPTR, TLSA, LOC, SSHFP, CAA and DHCID.

\begin{table}
\centering
\caption{Distribution of resource record type entries in the pDNS data set (Jun-Aug)}
\label{tab:distribution_rrtype}
\begin{tabular}{@{}lrr@{}}
\toprule
Type & \# Count   & Share   \\ \midrule
A      & 1,121,025,638 & 54.90\% \\
AAAA   & 197,388,865  & 9.67\%  \\
MX     & 682,948     & 0.03\%  \\
NS     & 7,662,147    & 0.38\%  \\
CNAME  & 156,708,021  & 7.68\%  \\
TXT    & 41,593,164   & 2.04\%  \\
NULL   & 432,232,574  & 21.17\% \\
Others & 84,371,709   & 4.13\%  \\ \bottomrule
\end{tabular}
\end{table}

\begin{figure}
  \centering
  \includegraphics[width=\linewidth]{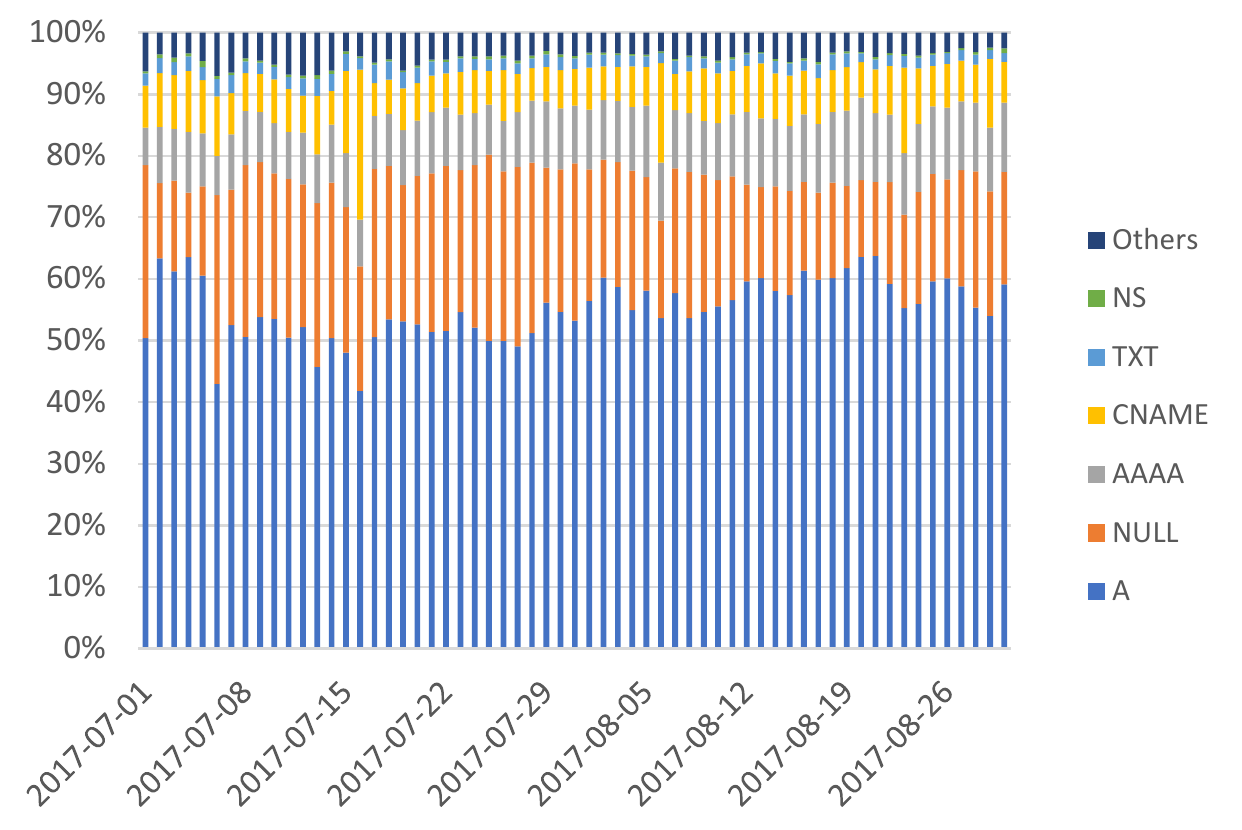}
  \caption{Percentage distribution of resource record types over time}
  \label{fig:rrtype_distribution}
\end{figure}

Figure~\ref{fig:rrtype_distribution} shows the distribution of the seen resource record types over time. 
The distributions remain stable over the measurement period and therefore demonstrate that we analyze a rather robust and representative data set.
NS type records and CNAME type records together represent a maximum of under ten percent per day.
Surprising is the large proportion of type NULL entries, such entries represent up to 30~\% of the total traffic every day. 
In total, about 21~\% are of type NULL.
According to RFC 1035 from 1987, this type is only experimental~\cite{rfc1035}. 
The rrtype TXT also make up a notable proportion of these entries.
All other rrtypes were rarely observed in the wild.

\subsubsection*{Distribution second-level domains to FQDNs}
\begin{figure}
  \includegraphics[width=\linewidth]{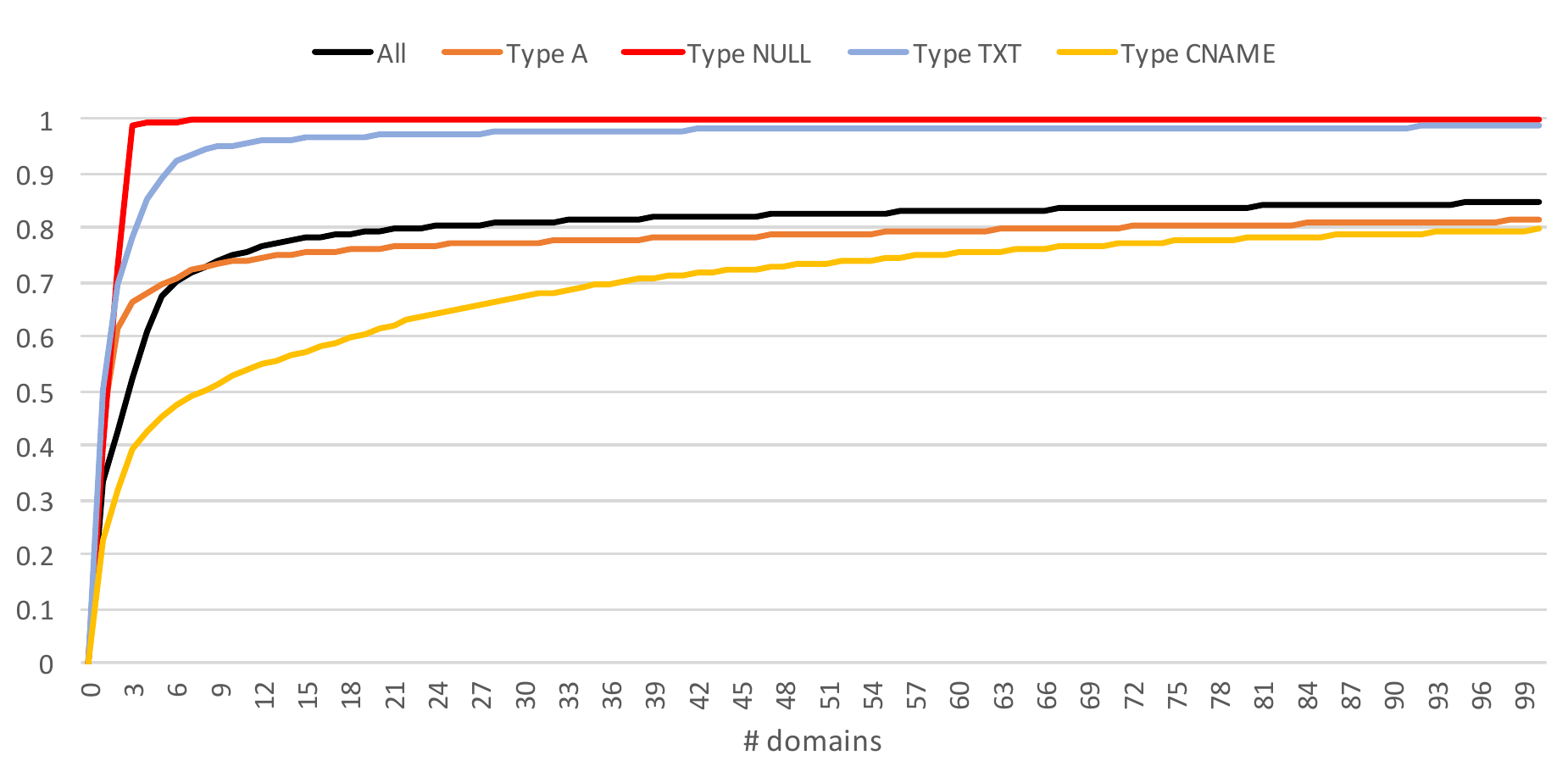}
  \caption{Cumulative distribution function for second-level domain names and their share of total FQDNs for all entries (black) and separated by record type A (orange), NULL (red), TXT (blue), and CNAME (yellow)}
  \label{fig:cdf_all}
\end{figure}

The most remarkable observation is that a small amount of second-level domain names are responsible for a large number of FQDNs. 
This insight implies that a few second-level domains generate a massive volume of subdomains.
Figure~\ref{fig:cdf_all} illustrates this behavior in a cumulative distribution function (CDF).
The black line represents all entries with all record types, whereas the orange one represents type A, the red line NULL, the blue line TXT, and the yellow line CNAME.
The CDF for all types (black line) indicates that roughly three second-level domain names are responsible for about 50~\% of the total newly observed FQDNs.
About 23 second-level domain names are responsible for 80~\%.
Additionally, we can see that the rise of the curve is slowly flattening, which means that many second-level domain names have very few new FQDNs.
The curve for record type A (orange line) is almost identical to the curve for all entries. 
It is noteworthy that especially type NULL and TXT contain even fewer domains, accounting for a major part of the total entries.
In contrast to this behavior, the curve of type CNAME is flattened, which shows a broader distribution of second-level domains to FQDNs.
We investigate these domains in more detail in Section~\ref{sec:study:details}.

\subsubsection*{Number of levels in the domain}
\begin{figure}
  \includegraphics[width=\linewidth]{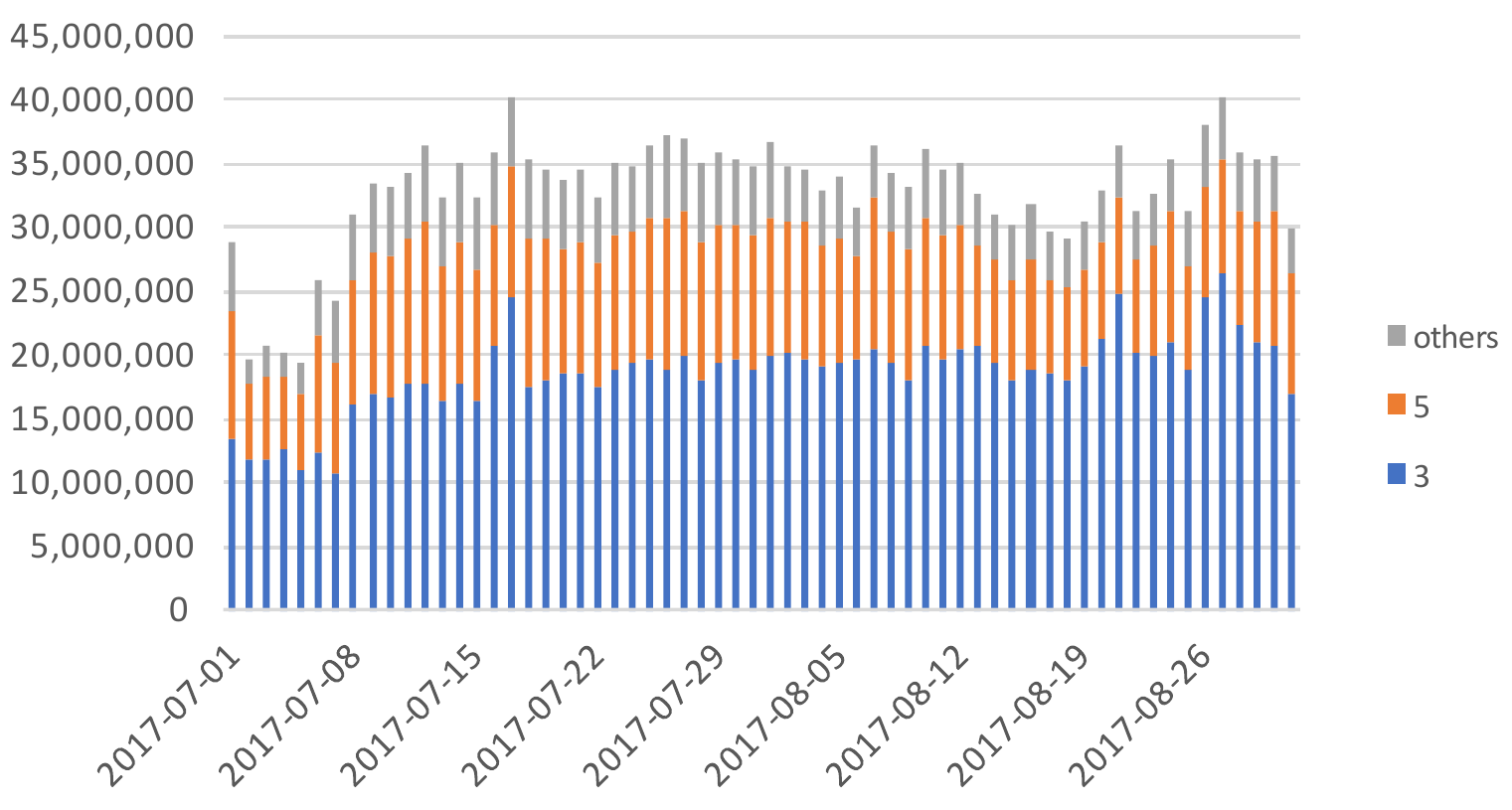}
  \caption{Distribution of FQDN levels over time}
  \label{fig:level_distribution}
\end{figure}

For the following evaluations, we count the TLD node as the first level.
The FQDN \textit{www.example.com.} therefore has level three.
As expected, more than 50~\% of the observed domains have a level three or lower in their rrname field.
These requests are usually simple IP address resolutions that occur when using the Internet and thus adapts to the rrtype distribution (type A $\sim$55\%).
Nevertheless, the proportion of on average level five or lower is about 30~\% and about 20~\% of the observed domains have more than five levels.  
This kind of queries are not common and may have different reasons (e.g., content delivery networks (CDNs) or DNS tunnel)
See Figure~\ref{fig:level_distribution} for the distribution of the number of levels in the FQDN over time.

\subsubsection*{Distribution rdata sizes}
\begin{figure}[t]
  \includegraphics[width=\linewidth]{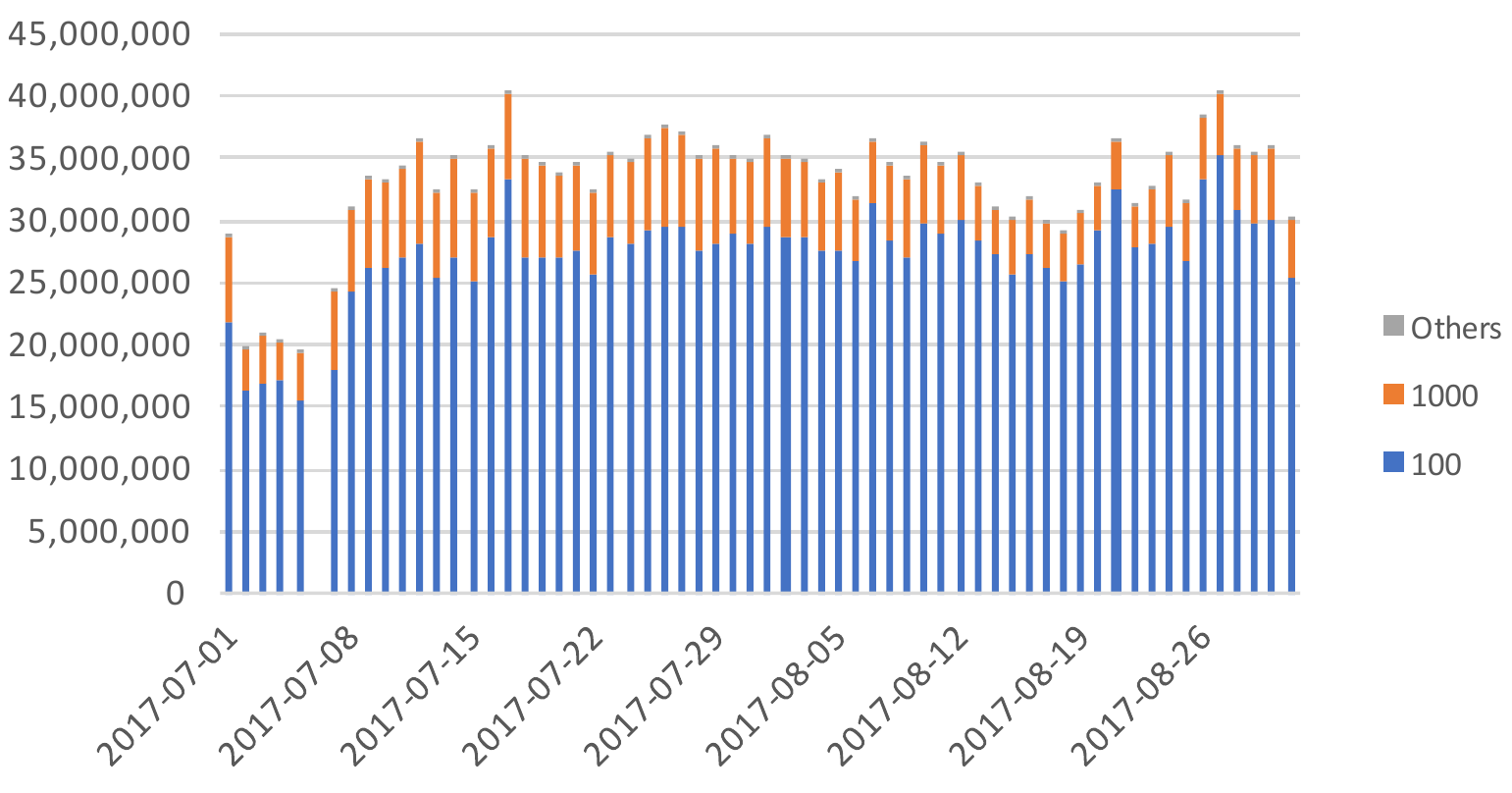}
  \caption{Distribution of rdata sizes over time}
  \label{fig:rdata_size_distribution}
\end{figure}

The rdata field represents the answer to the corresponding DNS query. 
Typically, the response should include an IP address, since DNS mainly translates domain names into IP addresses. 
The maximum size of 100 bytes including the dots and brackets should be sufficient for this purpose.
Nevertheless, a remarkable proportion contains more than 100 bytes. 
Figure~\ref{fig:rdata_size_distribution} shows this distribution over time.
The blue color represents all sizes up to 100 bytes (100), orange are the sizes up to 1000 bytes (1000) and grey are the sizes over 1000 bytes (Others).
Approximately 17.5~\% of rdata fields are between 100 and 1000 bytes in size.
A tiny proportion is even larger than 1000 bytes.

\begin{figure}
  \includegraphics[width=\linewidth]{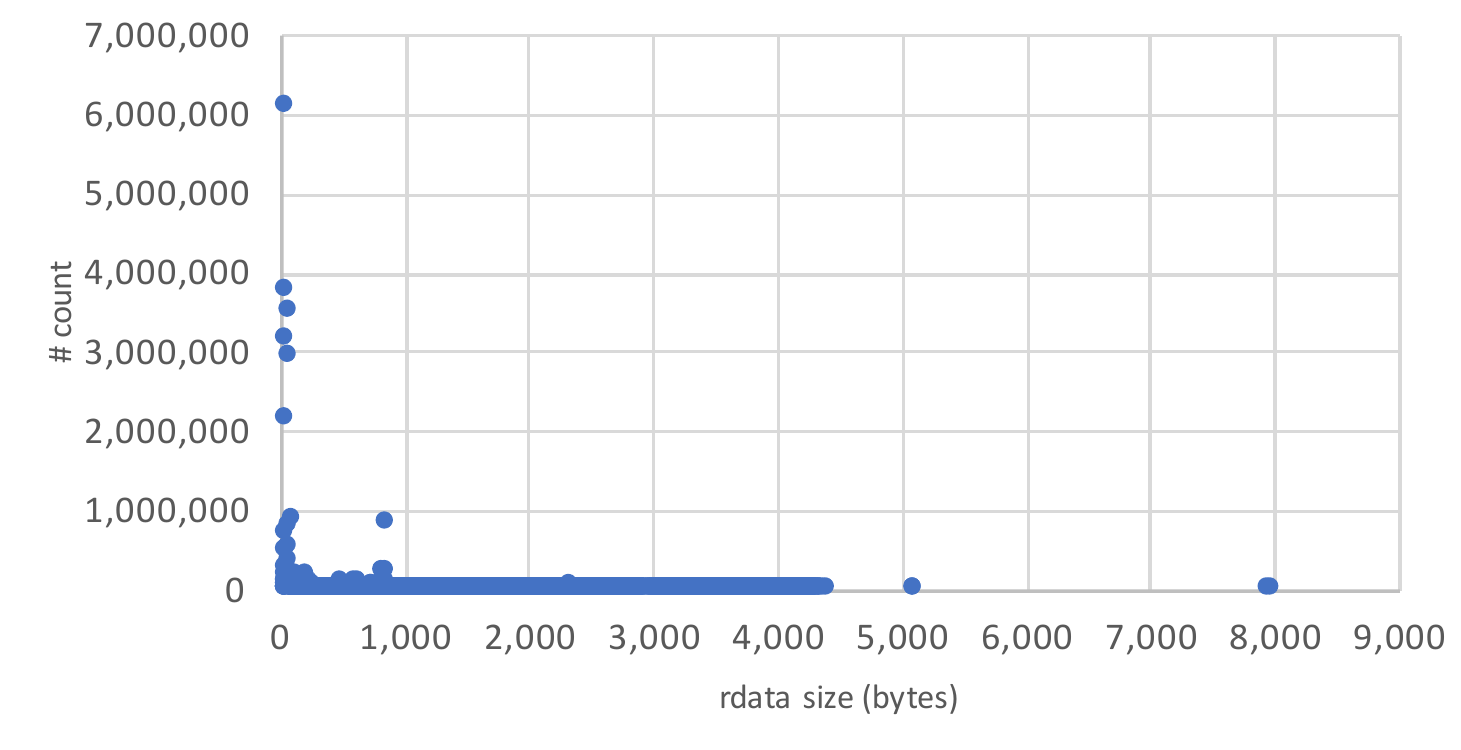}
  \caption{Scatter plot for distribution of the rdata size mean values over the whole data set}
  \label{fig:rdata_size_example}
\end{figure}

Figure~\ref{fig:rdata_size_example} shows a scatter plot with the distribution of the average rdata field sizes over the whole data set.
It is evident that a substantial part has tiny rdata sizes matching to domain to IP address resolutions since they include just IP addresses, but there are outliers on the X-axis, which have large rdata sizes.

\subsection{Additional Analyses and Results}
\label{sec:study:details}
Our results show a small number of domains is responsible for many DNS requests with new FQDNs. 
In the following, we explain the purpose of these domains.
Regarding the two months of our pDNS data set, the domains in Table~\ref{tab:top10_domains} are the top 10 domains with most DNS requests with new FQDNs.
The domain names \textit{ampproject.net}, \textit{53.de}, and \textit{spotilocal.com} alone represent more than 50~\% of the total traffic with new FQDNs in our data set. 
The domain name \textit{ampproject.net} belongs to the Google Accelerated Mobile Pages (AMP) project, which aims at accelerating access to mobile websites faster~\cite{ampproject}. 
The domain name \textit{53r.de} belongs to a German DNS tunnel provider, also the other three-character domain names \textit{8u6.de} and \textit{1yf.de} are part of it (see Section~\ref{sec:evaluation}).
Together, these three-character domain names make up a quarter of the total data.
The domain name \textit{spotilocal.com} in third place corresponds to the music streaming provider Spotify~\cite{npm_spotilocal}. 
The Spotify Desktop Client uses a web server running on localhost. 
The \textit{spotilocal.com} domain points to the Spotify localhost server and uses randomly generated subdomains to bypass browser limitations on the number of running concurrent connections to the same domain.
The fourth domain \textit{mts.ru} is related to the mobile provider Mobile TeleSystems in Russia.
The remaining domains are related to spyware (\textit{imrworldwide.com}), DNSSEC (\textit{dotnxdomain.net}), canary/decoy tools (\textit{cnr.io}), and a content management system (\textit{dynapsis.info}).

\begin{table}
\centering
\caption{Top 10 second-level domains with most pDNS entries}
\label{tab:top10_domains}
\begin{tabular}{@{}lrr@{}}
\toprule
Domain name      & Count     & Share   \\ \midrule
ampproject.net   & 681,017,564 & 33.37~\% \\
53r.de            & 192,389,690 & 9.43~\% \\
spotilocal.com   & 191,628,848 & 9.39~\% \\
8u6.de           & 185,147,960 & 9.07~\%  \\
1yf.de           & 125,973,029 & 6.17~\%  \\
mts.ru           & 52,553,371 & 2.58~\%  \\
imrworldwide.com & 35,496,798   & 1.74~\%  \\
dotnxdomain.net  & 23,290,118   & 1.14~\%  \\
cnr.io           & 20,820,485   & 1.02~\%  \\
dynapsis.info    & 19,784,924   & 0.97~\%  \\ \bottomrule
\end{tabular}
\end{table}

\begin{figure}[t]
  \includegraphics[width=\linewidth]{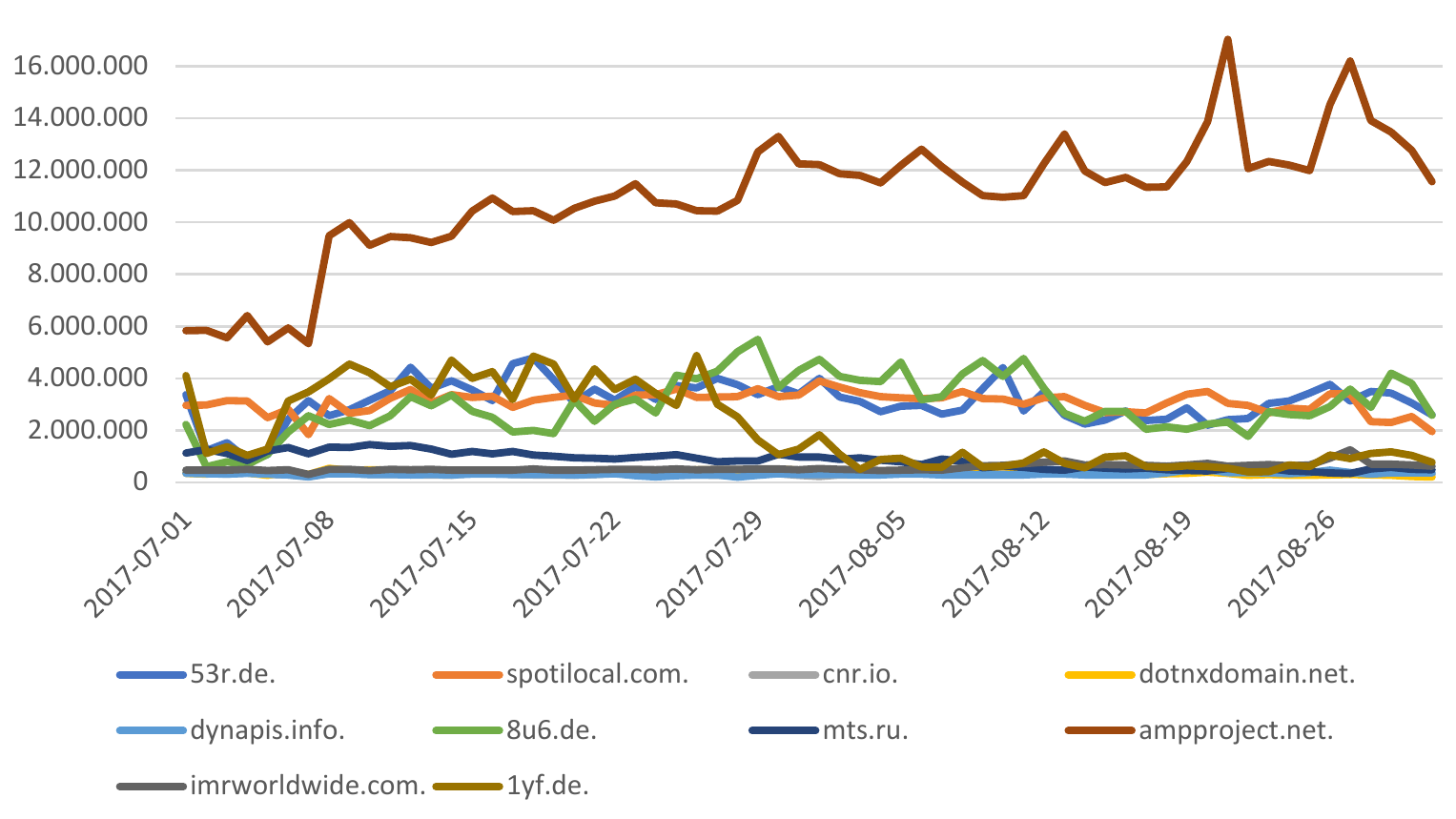}
  \caption{Behavior of the top 10 second-level domains with most pDNS entries over time}
  \label{fig:domains_over_time}
\end{figure}

An overview of the number of subdomains per domain for the top 10 domains over time is given in Figure~\ref{fig:domains_over_time}.
It illustrates that the amount of FQDNs per day remains rather stable for almost all top 10 second-level domains. 
There is an increase in \textit{ampproject.net}, which confirms that the AMP project is widely used and will likely be used more and more due to the increasing number of mobile devices. In addition, a linear regression analysis proves that the trend is significantly rising.
Furthermore, there was a descent at \textit{1yf.de} at the end of July and a continuous reduction at \textit{mts.ru}.

Most of the used resource record types among the top 10 domains with most pDNS entries are of type A (as expected).
Except for the three-character \textit{.de} domains, which do not use A resource records at all.
About 73~\% of all \textit{ampproject.net} entries are of type A and even all \textit{spotilocal.com} entries are of this type because these always resolve to localhost.
Comparing the second-level domains with the most entries between type A and AAAA, it is noticeable that \textit{ampproject.net} is responsible for most entries for both types.
In case of type A, this domain accounts for almost half of all queries and in case of type AAAA, it accounts for more than 92~\%.
Almost all requests for \textit{mts.ru} are of type A like almost all requests for \textit{dynapsis.info}.
Requests to \textit{imrworldwide.com} are of type CNAME.

Additionally, pDNS analysis might be used to measure the popularity of services.
An increase in requests for \textit{spotilocal.com}, \eg is an indication that Spotify is used more. 
Another example would be the use of Netflix using the domains \textit{netflixdnstest2.com} and \textit{netflixdnstest4.com} of type AAAA or \textit{nflxo.com} of type CNAME.
It is thus feasible to leverage pDNS to quantify the popularity from services that use DNS with new hostnames.
Note that the number of requested CDN domains in the domain field is negligible (less than 1~\% concerning the entire data set). 
Table~\ref{tab:top10_per_rrtype} contains information with the top ten second-level domains for resource record type A, AAAA, CNAME, and NS.

\begin{table*}
\centering
\caption{Top 10 second-level domains per type A, AAAA, CNAME, and NS}
\label{tab:top10_per_rrtype}
\begin{tabular}{@{}lrrllrr@{}}
\toprule
\multicolumn{3}{l}{Type A}               &  & \multicolumn{3}{l}{Type AAAA}            \\ \midrule
2nd level domain     & \# Count  & Share &  & 2nd level domain     & \# Count  & Share \\ \cmidrule(r){1-3} \cmidrule(l){5-7} 
ampproject.net.      & 499,006,190 &  44.1~\%     &  & ampproject.net.      & 182,009,815 &   91.7~\%    \\
spotilocal.com.      & 191,628,848 &  16.9~\%     &  & fbcdn.net.           & 4,083,343   &   2.1~\%    \\
mts.ru.              & 52,531,398  &  4.6~\%     &  & dotnxdomain.net.     & 4,030,857   &   2.0~\%    \\
dynapis.info.        & 19,174,477  &  4.1~\%     &  & ipv6test.com.        & 1,814,088   &   0.9~\%    \\
tekblue.net.         & 17,640,133  &  1.7~\%     &  & infernotions.com.    & 1,516,583   &   0.8~\%    \\
dotnxdomain.net.     & 15,582,070  &  1.4~\%     &  & ipleak.net.          & 970,214    &  0.5~\%     \\
beeline.ru.          & 12,279,081  &  1.1~\%     &  & dynapis.info.        & 610,447    &  0.3~\%     \\
netease.com.         & 6,959,973   &  0.6~\%     &  & ripe.net.            & 529,969    &  0.3~\%     \\
ipv6test.com.        & 6,947,460   &  0.6~\%     &  & netflixdnstest2.com. & 120,591    &  0.1~\%     \\
internetvikings.com. & 4,856,694   &  0.4~\%     &  & netflixdnstest4.com. & 120,050    &  0.1~\%     \\ \cmidrule(r){1-3} \cmidrule(l){5-7} 
                     &           &       &  &                      &           &       \\
\multicolumn{3}{l}{Type CNAME}           &  & \multicolumn{3}{l}{Type NS}              \\ \cmidrule(r){1-3} \cmidrule(l){5-7} 
2nd level domain     & \# Count  & Share &  & 2nd level domain     & \# Count  & Share \\ \cmidrule(r){1-3} \cmidrule(l){5-7} 
imrworldwide.com.    & 35,496,795  &  22.4~\%     &  & dotnxdomain.net.     & 495,525    &   6.3~\%    \\
igsonar.com.         & 14,239,271  &  9.0~\%     &  & dnsunions.net.       & 90,687     &   1.2~\%    \\
nflxso.net.          & 12,037,681  &  7.6~\%     &  & 154.in-addr.arpa.    & 88,902     &   1.1~\%    \\
1drv.com.            & 5,106,992   &  3.2~\%     &  & cu.cc.               & 78,532     &   1.0~\%    \\
sarahah.com.         & 3,914,675   &  2.5~\%     &  & qq.com.              & 62,411     &   0.8~\%    \\
lostmy.name.         & 3,862,838   &  2.4~\%     &  & superspeedcdn.com.   & 50,193     &   0.6~\%    \\
surveymonkey.com.    & 2,088,226   &  1.3~\%     &  & masimo.com.          & 42,904     &   0.5~\%    \\
messenger.com.       & 1,915,001   &  1.2~\%     &  & 23.in-addr.arpa.     & 40,072     &   0.5~\%    \\
seek.com.au.         & 1,888,242   &  1.2~\%     &  & goallurl.ru.         & 39,111     &   0.5~\%    \\
workable.com.        & 1,845,901   &  1.2~\%     &  & extrahop.com.        & 27,355     &   0.3~\%    \\ \bottomrule
\end{tabular}
\end{table*}

A more in-depth analysis of the record types NULL and TXT is worthwhile because these can transfer any data in their response field and thus fit for sending arbitrary information in a two-way communication channel well.
Additionally, previously known malware and tools use DNS tunnel with these particular types.

In summary, new hostnames are mainly used in three scenarios: usage by the Google AMP Project, utilization by Spotify, and DNS Tunnels. These three scenarios account for the majority of the collected data. Especially, the DNS tunnel aspect is interesting from a security perspective. Moreover, the large share of type NULL is unexpected.
\section{Identifying Suspicious Second Level Domain Names in Newly Observed Hostnames}
\label{sec:approach}

We have seen in the results of our empirical measurements that DNS tunnels are used a lot in practice. 
Next, we introduce an approach to identify potential suspicious domain names that may serve as DNS tunnels with the help of the information obtained from our measurement study.
First, we show the possibility to detect and further distinguish between multiple DNS tunnel setups in local networks and extract attributes we can use to find DNS tunnels in the pDNS data.
To make DNS tunnel detection efficient, we then introduce a filtering pipeline using the previously identified attributes and results of our conducted measurement study that reduces the size of the data set to simplify subsequent analyses. 

\subsection{Structural Analysis of DNS Tunnels}
\label{subsec:structure_analysis}

All DNS tunnel tools are easy to identify with an internal network view, e.g., monitoring the DNS resolver of a company network (see~\cite{qi2013bigram, homem2017harnessing, aiello2015dns, farnham2013detecting, dusi2009tunnel, sheridan2015detection, DBLP:journals/corr/abs-1004-4358}).
Tunnel tools significantly increase the number of requests (up to 2000\% more requests~\cite{van2008viability}), making identification often easy.
However, as the Internet has changed in recent years, this is no longer correct in all cases.
In particular, for the identification of DNS tunnels in aggregated global data, such as our Farsight data, the number of requests alone is not sufficient. 
It is not usable because nowadays there are many scenarios where many requests are sent to a second-level domain (e.g. see Section~\ref{sec:measurement} Google AMP, Spotify, or also CDNs).
For this reason, we need to find more attributes that can be used to identify DNS tunnels.

To first differentiate between DNS tunnel implementations, we built a test network.
In this network, we tested different DNS tunnel tools under laboratory conditions, generated traffic and saved it in PCAP files.
By analyzing the generated PCAPs, we were able to identify attributes that help to distinguish the individual DNS tunnel tools.
Therefore, we not only attempt to distinguish DNS tunnel traffic from regular DNS traffic, but also to determine the responsible DNS tunnel tool itself.

In our experiments, we used the following set of DNS tunnel tools: iodine~\cite{iodine}, dns2tcp~\cite{dns2tcp}, dnscat2~\cite{dnscat2}, dnscat~\cite{dnscat}, and OzymanDNS~\cite{ozymandns} as these are well-known DNS tunnel tools~\cite{merlo2011comparative,aiello2013performance}.
The tool iodine was first released in 2006 by Ekman and Andersson. It tunnels IPv4 packets through DNS and, thus, can be used for any protocol that runs on IPv4. It works on major Linux systems, Mac OS and Windows.
The tool dns2tcp was developed by Demvour and Collignon in 2008 and tunnels TCP traffic trough DNS.
The tool dnscat2 is the successor of dnscat that was released in 2004 as a Java based DNS tunneling tool.
OzymanDNS is a Perl tool developed by Kaminsky in 2004 for tunneling SSH over DNS.
Table~\ref{tab:tunnelimplementations} summarizes our utilized implementations with information about the latest commit of each tool and the supported resource record types.
We used iodine not only with the standard configuration (record type NULL) but also with type TXT, MX, SRV, CNAME, and A.
In comparison, DNS2tcp uses type TXT, dnscat2 utilizes three types (alternating CNAME, MX and NULL during operation), dnscat utilizes CNAME and OzymanDNS leverages type TXT.

\begin{table}
\centering
\label{tab:tunnelimplementations}
\caption{Utilized DNS Tunnel implementations}
\begin{tabular}{llll}
\toprule
tool      & latest commit & types & source                                                                                \\ \midrule
iodine    & 2018           & \begin{tabular}[c]{@{}l@{}}NULL,\\ PRIVATE,\\ TXT, SRV,\\ MX, A, \\ CNAME\end{tabular} & \cite{iodine}\\
dns2tcp   & 2017           & TXT                                                                                   & \cite{dns2tcp}\\
dnscat2   & 2015           & \begin{tabular}[c]{@{}l@{}}TXT, \\ CNAME, \\ MX\end{tabular}                                                                         & \cite{dnscat2}\\
dnscat    & 2005           & CNAME                                                                                   &  \cite{dnscat}\\
OzymanDNS & 2004           & TXT    & \cite{ozymandns}       \\ \bottomrule                                                                            
\end{tabular}
\end{table}

In addition to the DNS tunnel tools, we also tested two DNS tunnel providers, \textit{your-freedom.com} and \textit{tunnelguru.com}.
The difference between a DNS tunnel tool and a DNS tunnel provider is that the provider allocates the necessary infrastructure, which we have to set up with a DNS tunnel tool ourselves.
In our experiments, the two tunnel providers used the type NULL.

In total, we tested 12 different DNS tunnel implementations (five tools extended with iodine in five different configurations and two providers).
For each implementation, we extracted the most common values per attribute from our created PCAP files.
With this information, we created classes for each implementation in which possible values for the respective implementation are available.
During our analysis, the following eight attributes proved feasible to tell apart the DNS tunnel tools: 

\begin{enumerate}
    \item length of the FQDN without the third-level domain
    \item number of levels
    \item length of the fourth-level domain
    \item length of the fifth-level domain
    \item resource record type
    \item whether an encoding was used or not
    \item special characters at the beginning of the FQDN, and
    \item embedded particular substrings
\end{enumerate}

When a FQDN is mapped to the implementations, the values for the attributes are extracted from the examined domain and compared to each appropriate attribute per implementation. 

We tested in various experiments how many attributes should match in order to assume the examined FQDN to be associated comparable to the corresponding implementation. We defined that at least six matching attributes out of eight attributes represent similarly implementations.
With this simple method, we were able to assign 97\% of all seen DNS requests to the correct implementation in our generated data.

When using \textit{yourfreedom.com}, it is noticeable that a three-character \textit{.de} domain is always used (in our test cases \textit{53r.de}).
After further manual research we were able to assign other three-character \textit{.de} domains to \textit{yourfreedom.com} namely \textit{8u6.de}, \textit{1yf.de}, and \textit{2yf.de}~\cite{farsight_null}.
Through this experiment, we are now able to identify three-character \textit{.de} domains as DNS tunnel domains.
When using \textit{tunnelguru.com}, it is noticeable that a set of 53 three-character \textit{.in} domains are used in our experiments. 
The second-level domains randomly change per session through the set of TLDs. 
With this experiment, we can assign the seen three-character \textit{.in} domains to a DNS tunnel provider, too.

\subsection{Known Malware Utilizing DNS Tunneling}
\label{subsec:known_malware}

Besides the tools tested in our lab environment, we also analyzed a number of DNS tunnels that have been used in malware or by Advanced Persistent Threat (APT) groups. 
In the following, we provide a brief survey of the development of the use of DNS tunnels of malware.

In general, previously known DNS tunnel malware can be categorized by type of DNS usage, \ie C2 communication or data exfiltration. 
In addition, it is also possible to group them according to the type of malware or type of attack target. 
There is malware for payment terminals, malware for bot distribution and control, and malware for targeted network attacks.
Overall, it is noteworthy that the examples of malware described in the following always use type NULL or TXT.
The utilization of both types is reasonable, as text data or even any kind of data may be transmitted in the response. 
Thus, a two-way communication is ideal to implement.

In August 2011, attention was drawn to the Morto worm~\cite{morto}, which used a DNS tunnel for C2 communication.
In the same year in September, another malware was analyzed (Feederbot)~\cite{dietrich2011botnets}. 
Feederbot is a botnet malware that also uses DNS as a C2 communication channel.
At the beginning of 2014, a remote access Trojan appeared (PlugX Variants)~\cite{plugx}, of which a module implements C2 communication via DNS. 
In October 2014, a malware (FrameworkPOS) was discovered that implements data exfiltration using DNS requests~\cite{frameworkpos}. 
It targets Point of Sale (POS) systems.
Another POS malware that also uses DNS as an exfiltration channel is BernhardPOS from November 2015~\cite{bernhardpos}.
In 2016, there were several malware samples and APT groups that used DNS tunnels for their purposes~\cite{multigrain,C3PRO-RACCOON,apt34,wekby,oilrig}.
Multigrain is a POS malware using DNS as exfiltration channel~\cite{multigrain}. 
C3PRO-RACCOON used DNS tunnels for establishing a C2 communication channel during a botnet campaign~\cite{C3PRO-RACCOON}.
The APT groups APT34 and Wekby both make use of DNS tunnels for C2 communication~\cite{apt34,wekby}.
In particular, the Oilrig campaign in May and October used DNS as a communication channel by the malware Helminth and ISMAgent~\cite{oilrig}.
This campaign is loosely aligned with APT34.
The Remote Access Trojan DNSMessenger was discovered in March 2017~\cite{dnsmessenger}.
Moreover, in 2017, the APT32 group and another malware (Alma Communicator) from Oilrig became noticeable~\cite{apt32,oilrig_alma_communicator}. 
The latest malware in 2018 is UDPoS and it is a POS malware that uses DNS for data exfiltration~\cite{udpos, udpos_first}.

\subsection{Filtering approach}
\label{subsec:approach}

We use some of these common and uncommon attributes for the reduction of our data and spotting potential DNS tunnel domain names in course of a step-wise filtering. More specifically, the lessons learned from Sections~\ref{sec:measurement}, \ref{subsec:structure_analysis} and~\ref{subsec:known_malware} help us to develop a filter approach for the aggregated pDNS data for a second measurement study on the use of DNS tunnels in the wild.
The following filter functions are carefully created manually based on our insights.

\begin{enumerate}
    \item Number of subdomains.
    \item Level of full domain (FQDN). 
    \item Resource record type (rrtype).
    \item Size of response (rdata). 
    \item Known non-DNS tunnel use cases \eg DNS-based mail authentication or reverse DNS lookups. 
    \item Known second-level domains.
    \item Entropy.
	\item Character or bigram frequency~\cite{qi2013bigram}. 
\end{enumerate}

Note that we do not need an in-depth analysis of the responses since we want to detect besides two-way communication channels (up and downstream) also upstream-only channels which do not need any responses (6).
In addition, we do not use the already known attributes entropy (7) and character or bigram frequencies (8) but focus on the information that we can obtain directly from the pDNS data without any further processing.
With five features, we developed our step-wise filtering approach for the identification of DNS tunnels.
The resource record type is an excellent prefiltering attribute since our observations during the measurement study verify it as a good starting point to effectively reduce a large amount of pDNS data. The level of the FQDN, the number of subdomains per second-level domain, known non-DNS tunnel use cases, and known second-level domains are further attributes we utilize to reduce the data set.

It is worth noting that we do not refer to local data with information per client but to pDNS data, which allows us to make statements on the global usage of new FQDNs.
In local networks, the number of subdomains to a second-level domain is usually enough to detect a DNS tunnel. With our aggregated data this does not work anymore because, otherwise, we would get a lot of false positives~\cite{plight}. However, the combination of five filter functions allows us to identify DNS tunnels even in aggregated data.

\begin{figure*}[ht]
\centering
	\includegraphics[width=.8\linewidth]{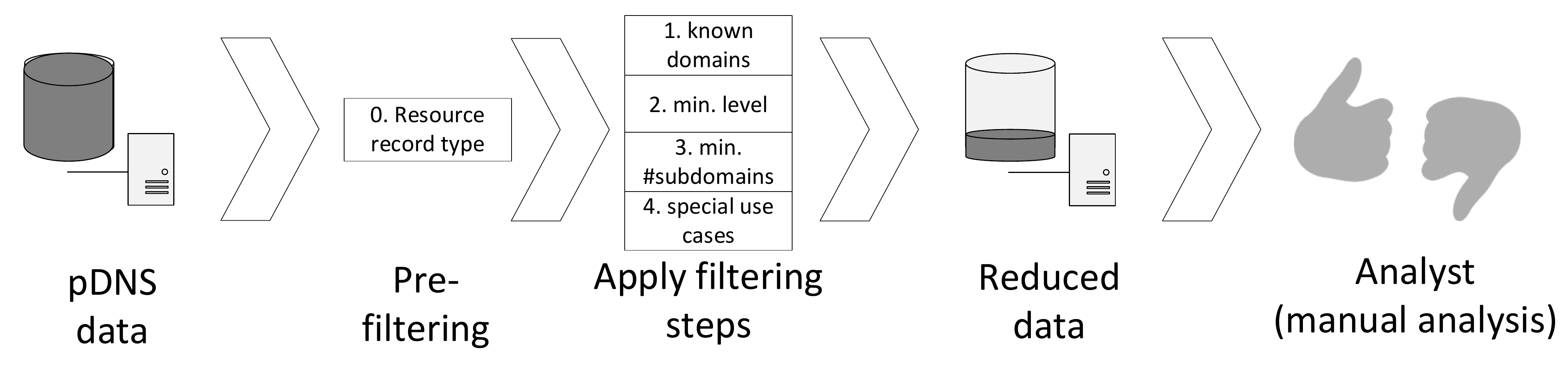}
	\caption{Workflow of our filtering of pDNS data. As input we use our gathered pDNS data. Next, our filter pipeline begins with prefiltering regarding resource record type. Afterwards, we apply further filter steps (known domains, minumum level, minimum number of subdomains per second-level domain, special use cases). Consequently, we obtain a reduced data set which can then be manually examined by an analyst to decide whether it is a possible tunnel domain or not.}
	\label{fig:workflow}
\end{figure*}

Figure~\ref{fig:workflow} illustrates the approach for filtering our gathered pDNS data for potential suspicious DNS tunnel domains.
As input, we use the gathered pDNS data from Farsight SIE. 
During our measurement study in Section~\ref{sec:measurement} and the DNS tunnel survey that we conducted in Section~\ref{subsec:structure_analysis} and Section~\ref{subsec:known_malware}, we discovered that tunnels predominantly use type NULL or TXT. Therefore, in a first step, we prefilter for the corresponding resource record types (0). The other types, like A or AAAA, are filtered out because we found that in theory tunnels can be implemented with these types but in practice, they are not used.
DNS tunnels utilize types like NULL or TXT because these can transmit arbitrary information.
After prefiltering, further filter functions are applied. We begin with filtering functions that remove as much as possible at the very beginning in order to make the following filtering and analyses execute on small data sets. It is necessary that the steps are all performed to ensure proper results. Note that for a detailed analysis, the results of the individual filter steps may be saved, analyzed, and more customized.

We start by filtering known second-level domains (1). For the known domains, we use a list of known CDN domains and already known tunnel domains.

After that, we filter for requested FQDNs with at least level four (2).
We observed that DNS tunnel FQDNs usually include a short constant third-level domain below the second-level domain. A third-level domain is mostly required as it is elaborate to get an authoritative DNS server for second-level domains. However, theoretically tunnels could also use second-level domains for transferring information. The tunnels we investigated avoid this extra effort (i.e., running a DNS server) by merely using a third-level domain. Additionally, the third-level domain is short because the FQDN should encode as much information as possible since the length of an FQDN and the length of its subdomains itself are limited.

Next, we filter for at least two different subdomains per second-level domain (3). When using DNS tunnels for data transmission the straightforward way is to embed data inside the requested FQDN. Thus, a certain number of subdomains must always be generated, since most data transfers are larger than the data size that can be sent in a single DNS request. In other words, data must be split into smaller packets and then processed with many DNS requests in order to transmit it via DNS tunnels.

In the last step (4), we filter for known non-DNS tunnel usage patterns. For example, the \textit{.arpa} TLD entries, which are used exclusively for infrastructure purposes~\cite{rfc1035}, or DNS-based mail authentication mechanisms such as the Sender Policy Framework (SPF)~\cite{rfc7208} or DomainKeys~\cite{rfc6376}. 

After using our step-by-step filter approach, a reduced data set with potentially suspicious candidate domains remains. This should then be checked manually by an analyst to finally decide whether the domain candidates are DNS tunnel domains or not.

\section{Filtering Results}
\label{sec:evaluation}

In the following, we present the results of our filtering approach to identify potential DNS tunnel domains. 
First, we test our step-wise filtering method by checking each filter function on self-generated traffic.
Next, we discuss the results by prefiltering resource record type NULL and TXT since these types are used by well-known malware (see Section~\ref{subsec:known_malware}), and six out of seven tested tools or providers support them too (see Section~\ref{subsec:structure_analysis}).
To complete this, we consider results for prefiltering by other types, e.g., CNAME.

\subsection{Filter Function Evaluation}
First, we demonstrate that our individual filtering steps are applicable and do not filter known DNS tunnel packets.
We examine the individual filter functions on self-generated DNS tunnel data. In addition, we analyze DNS data from a standard computer for about one month (12th Oct 2018 to 9th Nov 2018).
We did not observe a single type NULL or TXT packet. 
In the reverse experiment, we confirmed that the tunnel implementations mostly utilize types NULL or TXT.
The subsequent filtering for at least level four also does not reduce any further entries from our tunnel PCAPs, as well as filtering for at least two subdomains per second-level domain.
This, therefore, confirms that our filter sequence does not remove any potential DNS tunnel domains from our tested implementations.

\subsection{NULL Domains}
The results of the prefiltering by type NULL show that a total of almost 99 percent of this type represents DNS tunnel traffic (three-character \textit{.de}, see Section~\ref{subsec:structure_analysis}).
By applying our further filter steps, 96 potential DNS tunnel domains remain.
According to manual analysis, these domains contain four domains that are related to an APT campaign (APT 32). The remaining domains are potential DNS tunnel domains, 35 of which are unknown so far. About 80 percent of these domains are iodine-like domains.

\subsubsection{Detailed filtering steps}
\subsubsection*{Prefiltering}
After filtering by record type NULL, we reduced the whole data set by more than 70~\% so that 96 different second-level domains with 439,463,986 pDNS entries remain as our NULL domains.
\subsubsection*{Known domains}
The filtered data set still contains three-character \textit{.de} domains (\textit{53r.de}, \textit{8u6.de}, \textit{1yf.de}, and \textit{2yf.de}), which all belong to a DNS tunnel provider as learned in Section~\ref{subsec:structure_analysis}).
The share of these four second-level domains is more than 99~\%, \ie almost the entire type NULL traffic can be classified as DNS tunnel traffic.
The second most common category of domains are three-character \textit{.in} domains, which we can also connect to a DNS tunnel provider (again see Section~\ref{subsec:structure_analysis}).
The total amount of three-character \textit{.in} domains is 1,120,114 (0,25\%) and belongs to 52 second-level domains (\eg \textit{qv4.in}, \textit{mm4.in}, \textit{na2.in}, etc.).
Therefore, we conclude that NULL traffic should always be paid special attention to. 
\subsubsection*{Min. level and min. \#subdomains}
We filtered by level (min. level four) and by the number of subdomains (min. two FQDN per second-level domain), which did not remove any further entries.
\subsubsection*{Special use cases}
With NULL traffic, we did not consider any special use cases and therefore did not perform further filtering in this step.
\subsubsection*{Manual analysis}
In the following, we discuss more information about the remaining domains.
Therefore, we remove the known three-character domains resulting in 40 second-level domain candidates with 1,900,389 pDNS entries for other possible DNS tunnels.
After a manually performed Google web search, we were able to identify \textit{dashnxdomain.net} as a non-DNS tunnel domain.
The remaining 39 second-level domains showed typical DNS tunnel behavior, including a lot of requests in a short period of time and randomly generated subdomains.
Furthermore, we identified four domain names (\textit{gl-appspot.org}, \textit{facebook-cdn.net}, \textit{tonholding.com}, and \textit{nsquery.net}) that were used by APT group APT32 (see Section~\ref{apt32} for details).

\begin{table*}[htb]
\centering
\caption{Remaining second-level domains after filtering by Type NULL and known domains}
\label{tab:remaining_null}
\begin{tabular}{@{}llrrrr@{}}
\toprule
second-level domain                         &
bailiwick                                & \# FQDNs & \# days &
iodine-like & group \\ \midrule
dicksin.me.                              &
sub.dicksin.me                           & 1,044,739  & 5            &   
y      &      Others       \\
toc.sc.                                  &
de.toc.sc                                & 387,435   & 36           &   
y      &     Service        \\
daemonslayer.net.                        &
tunnel.daemonslayer.net                  & 103,081   & 2            &  
y       &      Service       \\
2cb262aa-...-4b772a5ee2df.ca.          &
2cb262aa-...-4b772a5ee2df.ca.          & 74,261                  &
 3            &    n       &      Others       \\
%2cb262aa-cbb3-4e69-8f66-4b772a5ee2df.ca. &
%%2cb262aa-cbb3-4e69-8f66-4b772a5ee2df.ca. & 74261               &
%3           &  n         &        Others     \\
 ro.lt.                                   & tunz.ro.lt                               & 62,740    & 1            &    y       &      Others       \\
uk.to.                                   &
tunz.uk.to                               & 42,398    & 2            &  
y       &     Service        \\
itsaunixsystem.net.                      &
t.itsaunixsystem.net                     & 36,006    & 1            &   
y      &      Private       \\
mooo.com.                                &
purple-cow.mooo.com                      & 18,177    & 9            &
n    &                Service       \\
mst-pro.ru.                              &
d2.mst-pro.ru                            & 14,080    & 4            &   
y      &      Organization       \\
opusbit.com.                             &
i.opusbit.com                            & 10,956    & 3            &   
y      &     Service        \\
uux1.com.                                &
uu.uux1.com                              & 8,353     & 5            &
n    &                  Others      \\
dillonbeliveau.com.                      &
t1.dillionbeliveau.com                   & 8,097     & 1            &   
y      &      Private       \\
cehturkiye.com.                          &
vpn.cehturkiye.com                       & 5,048     & 3            &   
y      &      Service       \\
azvw.org.                                &
io.azvw.org                              & 4,888     & 1            &   
y      &      Service       \\
fajri.info.                              &
fajri.info                               & 4,877     & 1            &
n    &                 Private      \\
cokeduptrading.com.                      &
iodinens.cokeduptrading.com              & 4,067     & 2            &   
y      &      Others       \\
ethicalreporting.org.                    &
tunnel.ethicalreporting.org              & 3,148     & 1            &   
y      &      Organization       \\
insmedportal.com.                        &
t.insmedportal.com                       & 3,037     & 1            &   
y      &      Others       \\
allconnect.com.                          &
metuchen.allconnect.com                  & 3,014     & 2            &   
y      &      Service       \\
clubarsenal.ru.                          &
home.clubarsenal.ru                      & 2,975     & 3            &   
y      &      Private       \\
ab0.tj.                                  &
io.ab0.tj                                & 2,863     & 1            &   
y      &      Others       \\
zensecurity.su.                          &
d.zensecurity.su                         & 2,352     & 1            &   
y      &    Organization         \\
pwnintended.com.                         &
t1ns.pwnintended.com.                    & 754      & 1            &   
y      &     Others        \\
vorner.cz.                               &
dnsvpn.vorner.cz                         & 718      & 1            &   
y      &     Private        \\
zestysoft.com.                           &
t1.zestysoft.com                         & 408      & 1            &   
y      &     Private        \\
us.to.                                   &
blipi.us.to                              & 209      & 2            &   
y      &     Others        \\
x86sec.com.                              &
iodine.x86sec.com                        & 100      & 1            &   
y      &      Service       \\
vasi.li.                                 &
t.vasi.li                                & 94       & 2            &  
y       &     Private        \\
khashaev.ru.                             &
ns.khashaev.ru                           & 22       & 1            &  
y       &      Private       \\
getgaze.com.                             &
i2.getgaze.com                           & 17       & 2            &
n    &                Others        \\
ambrisko.com.                            &
tunnel2.ambrisko.com                     & 4        & 1            &
n    &               Private        \\
notf2pool.com.                           &
d.notf2pool.com                          & 4        & 2           &
n    &                 Others       \\
thegnet.tk.                              &
t1.thegnet.tk                            & 4        & 1           &
n    &              Private         \\
plak.cc.                                 &
t.t.plak.cc                              & 3        & 1           &
n    &               Others         \\
bgasecurity.com.                        &
tunnel.bgasecurity.com                   & 3        & 1           &
n    &                Organization        \\ \bottomrule
\end{tabular}

\end{table*}

Finally, we analyzed the remaining 35 domains in more details (see Table~\ref{tab:remaining_null}). 
Initially, we categorized these domains based on Google search results into four groups. 
We differentiated between the group Service-related ($\sim$26\%), such as Web applications or blogs, the group Organization-related ($\sim$11\%) which includes companies, the group Private-related ($\sim$29\%), wherein individuals use tunnels, and the group Others ($\sim$34\%), in which it became difficult to find a precise explanation.
We investigated that the third-level domain \textit{tunnel} or \textit{tunnel2} is used by four second-level domains (\textit{daemonslayer.net}, \textit{ethicalreporting.org}, \textit{ambrisiko.com}, and \textit{bgasecurity.com}).
The third-level domains \textit{t}, \textit{t1}, and \textit{t1ns} are used by eight second-level domains out of the groups Private and Others.
And the third-level domains \textit{tunz}, \textit{iodine}, \textit{iodinens} are used by four second-level domains.
All these domains give a clear sign for being tunnel domains since their third-level domains are rather short and refer textual to tunnels.
For the remaining 19 domains, we do not have an extra indicator for DNS tunnel usage, but the behavior is, in any case, DNS tunnel comparable.

As a final step, we assigned the remaining domain names to our tested DNS tunnel implementations.
We could match almost 80\% of the traffic to iodine in the default settings.

\subsection{TXT Domains}
The results of prefiltering by type TXT show that a total of about 35 percent of this type represents DNS tunnel traffic.
By applying our further automated filter steps, 233 potential DNS tunnel domains remain.
According to manual analysis, these domains contain different domains that are related to companies, universities, video streaming, and potential DNS tunnel domains.
Finally, we found another APT campaign (Wekby) here as well. 

\subsubsection{Detailed filtering steps}
\subsubsection*{Prefiltering}
Filtering type TXT reduces the data set by more than 97~\% so that 175.852 second-level domains with 42.175.478 pDNS entries remain.
In this case, considerably more second-level domains with far less FQDNs compared to the NULL domains.
\subsubsection*{Known domains}
The second-level domain with most subdomains is \textit{cnr.io}. 
This domain belongs to Canary Tools (by Thinkst Applied Research) and can, therefore, be filtered. 
This step allows us to remove more than 21 million entries and further halve the reduced data set.
Furthermore, it is conspicuous that three-character \textit{.de} domains are prevalent here, too (14,971,251 entries).
About 35~\% of the domains in our TXT domains are related to DNS tunnels because we already know them as DNS tunnel domains and can filter them accordingly (see Section~\ref{subsec:structure_analysis}).
\subsubsection*{Min. level and min. \#subdomains}
Since the number of potential domains presumably related to DNS tunnels, which we refer to as domain candidates is still high, they cannot be validated by hand. 
We use further filtering based on the level (min. level four) and at least two FQDNs per second-level domain. 
These filter steps reduce the data set to 7,700 potential DNS tunnel candidates. 
\subsubsection*{Special use cases}
In the next step, we reduced known non-DNS tunnel use cases from the data set. 
We removed DNS mail authentication mechanisms (\eg SPF, DKIM, DMARC, DomainKeys) and rDNS requests.
Due to the different filtering, it was possible to reduce the number of potential domains to 233.
\subsubsection*{Manual analysis}
When looking at the domains with most FQDNs, it is noticeable that some companies and universities appear, for example \textit{arcticwolf.net}, \textit{extrahop.com}, \textit{berkeley.edu}, or \textit{nlnetlabs.nl}. 
The domain with most of the pDNS entries is \textit{ksx.la} and seems to be related to the domain \textit{knb.la} because the structure of the FQDNs (length, level, randomization) and the behavior (many subdomains) is identical for both second-level domains.
We have a total of 15 second-level domains seen every day. 
In these domains we recognize the following five groups. 
\begin{enumerate}
\item company domain names such as \textit{arcticwolf.net}, \textit{extrahop.com}, or \textit{brightmail.com}. (A total of six domains can be counted as company domains)
\item video streaming domains, \ie \textit{erlyvideo.org}.
\item universities or nonprofit organizations such as \textit{berkeley.edu} or \textit{nlnetlabs.nl}. 
\item the domains \textit{dsipsl.net}, \textit{dsomc.net}, \textit{dsoml.net}, and \textit{dsrmc.net} seem to belong together by structure.
\item other domains that we can not assign (\textit{pf-d.ca} and \textit{ymapp.com}).
\end{enumerate}

Next we filter the daily seen second-level domains. 
After filtering these, there are 216 domain candidates left to check.
The domains which only have one entry left after filtering are removable as it is not possible to create a useful tunnel.
This led us to 156 domain candidates. 
A further correlation with the Alexa top one million domains allows reducing the number of domain candidates by another 28 domains since we assume the Alexa top one million domains are not used for DNS tunnels.
The remaining 128 domains are suspicious tunnel domains.

Some second-level domains still belong to companies \eg \textit{allconnect.com}, \textit{safedns.com}, \textit{panorama9.com}, and \textit{eset.com}.
However, we also find suspicious domains like \textit{engineershow.com}, \textit{sharepoint-microsoft.co} or \textit{newsfeeds-microsoft.press}. 
\textit{engineershow.com} seems to be used by a malware and the other two domains are IOCs of the group Copy Kitten~\cite{wiltedtulip}. 
Also interesting is the domain \textit{wetun.nl}, which was used for a CTF where iodine traffic had to be analyzed~\cite{ctf_wetun}.
Among the remaining domains, it was possible to find even more indicator of compromises (IOCs) used by the Wekby APT group to tunnel data via DNS.
We provide further details in Section~\ref{wekby}.

\subsection{Other Resource Record Types}
For the remaining types it is worth taking a closer look at CNAME, as three of the five tested tools in Section~\ref{subsec:structure_analysis} also support CNAME.

After prefiltering by type CNAME and applying our further filter steps, 182,205 potential DNS tunnel domains remain.
This number of candidate domains is too large to be fully inspected manually. Therefore, we examined the Top 100, which account for almost 80 percent of all entries.
These domains contain domains related to companies, universities, video streaming, and potential DNS tunnel domains.
Nevertheless, we were unfortunately not able to identify any other previously unknown DNS tunnel domains.

We did not take a closer look at the other types, as they are not commonly used for DNS tunnels (as we discovered in Section~\ref{sec:approach}).

\section{Case Studies}
\label{sec:casestudies}

After identifying potential DNS tunnels in our data set, we present two case studies about the utilization of DNS tunnels used in Advanced Persistent Threat (APT) campaigns. 
An APT is usually a targeted network attack in which unauthorized persons gain access to a network and remain undetected as long as possible~\cite{daly2009advanced}. 
Targets are often organizations with valuable information, \eg governments, manufacturers, or the financial sector.
The case studies confirm that even malicious DNS tunnels are found through our approach and that this is a real-world threat. Note, already known DNS tunnels are the only way to show that our approach works since we do not have ground truth data. However, the other potential DNS tunnel domains detected in Section~\ref{sec:evaluation} are new and unknown.

\subsection{APT 32}
\label{apt32}
\begin{figure}
\centering
  \includegraphics[width=\linewidth]{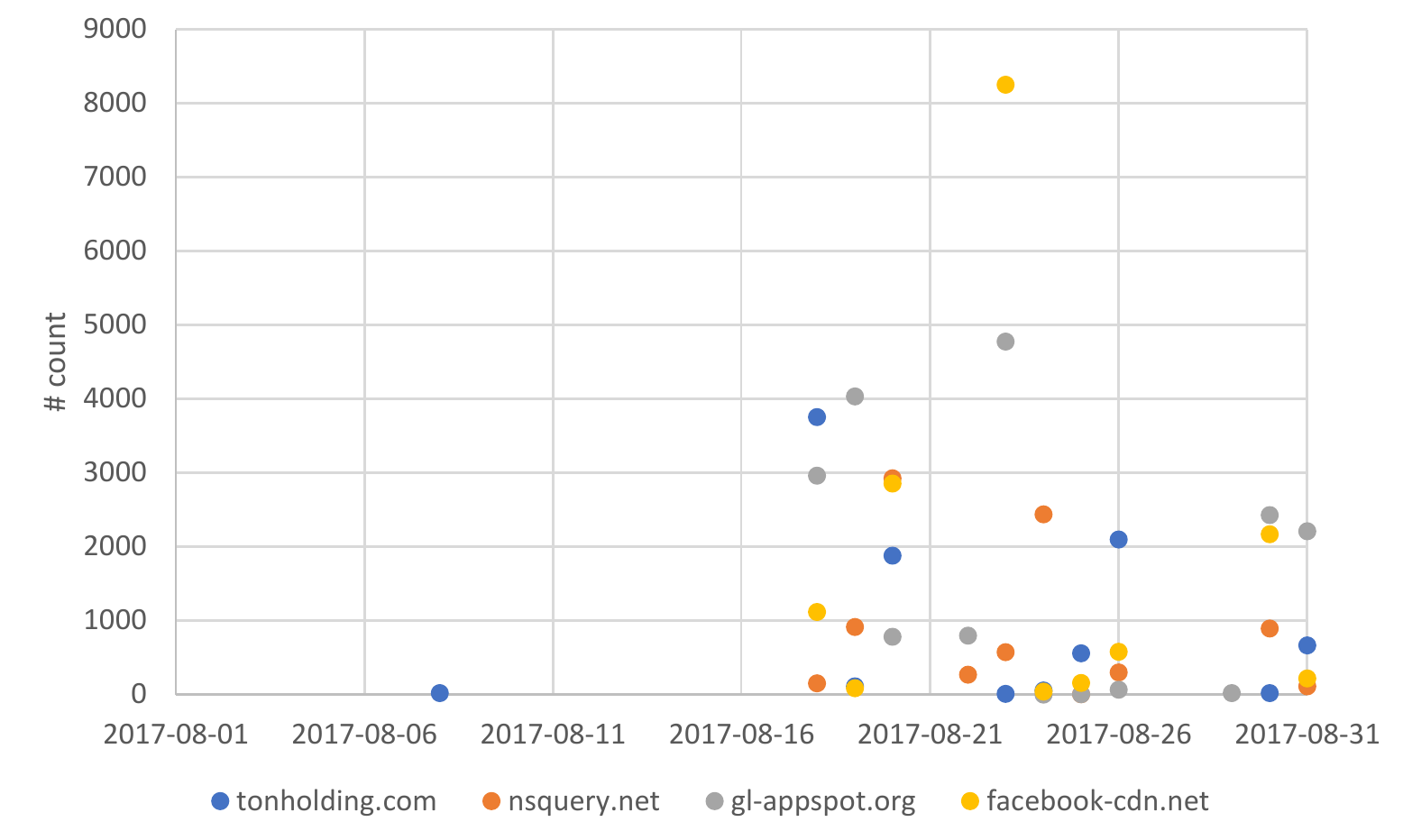}
  \caption{Distribution of identified malicious APT32 domains over time (August 2017)}
  \label{fig:aptdomain_distribution}
\end{figure}

APT 32 (\emph{OceanLotus Group}) is an APT group that was uncovered in mid-May 2017~\cite{apt32}.
Through our filter approach for identifying potential DNS tunnels, we identified four domains used by APT 32 in August 2017 with type NULL.
Figure~\ref{fig:aptdomain_distribution} summarizes the occurrence of these second-level domains per day.
It is visible that the second-level domains found by the introduced pDNS filtering approach (\textit{tonholding.com}, \textit{nsquery.net}, \textit{gl-appspot.org}, and \textit{facebook-cdn.net}) are used for data transmission (a large number of subdomains). 
The first substantial use took place between August 18th and August 20th, \ie a weekend (Friday to Sunday).
The day with the most requests was August 23rd (Wednesday).
There is no clear pattern at the time of use.
A more extended analysis period might be interesting to identify patterns in its usage.
The most requested domain during our records related to APT 32 is \textit{gl-appspot.org} (18.781 queries).
However, \textit{facebook-cdn.net} (15.504 queries) seems to be important for the infrastructure of APT 32 since it is used as email domain in the SOA records of all other DNS tunnel APT 32 related domains.

Through further research, we searched for all known indicator of compromises (IOCs)~\cite{apt32} for APT 32 in our data set. 
We were thus able to find three more second-level domains which were used during our data collection time (\textit{shalaghlagh.tk}, \textit{teriava.com}, \textit{ntpudateserver.com}).
The domain \textit{teriava.com} seems to represent a keepalive bit, as this domain was periodically resolved once every other day during our measuring period. 
The other domains were barely noticed. \textit{ntpudateserver.com} was spotted twice once with 19 entries on 07/17/2017 and once with four entries on 08/11/2017, and \textit{shalaghlagh.tk} only once on 7/7/17 with two entries. We conclude that these domains are not used for data transfer but probably implement C\&C communication.

\subsection{Wekby}
\label{wekby}

\begin{figure}
\centering
  \includegraphics[width=\linewidth]{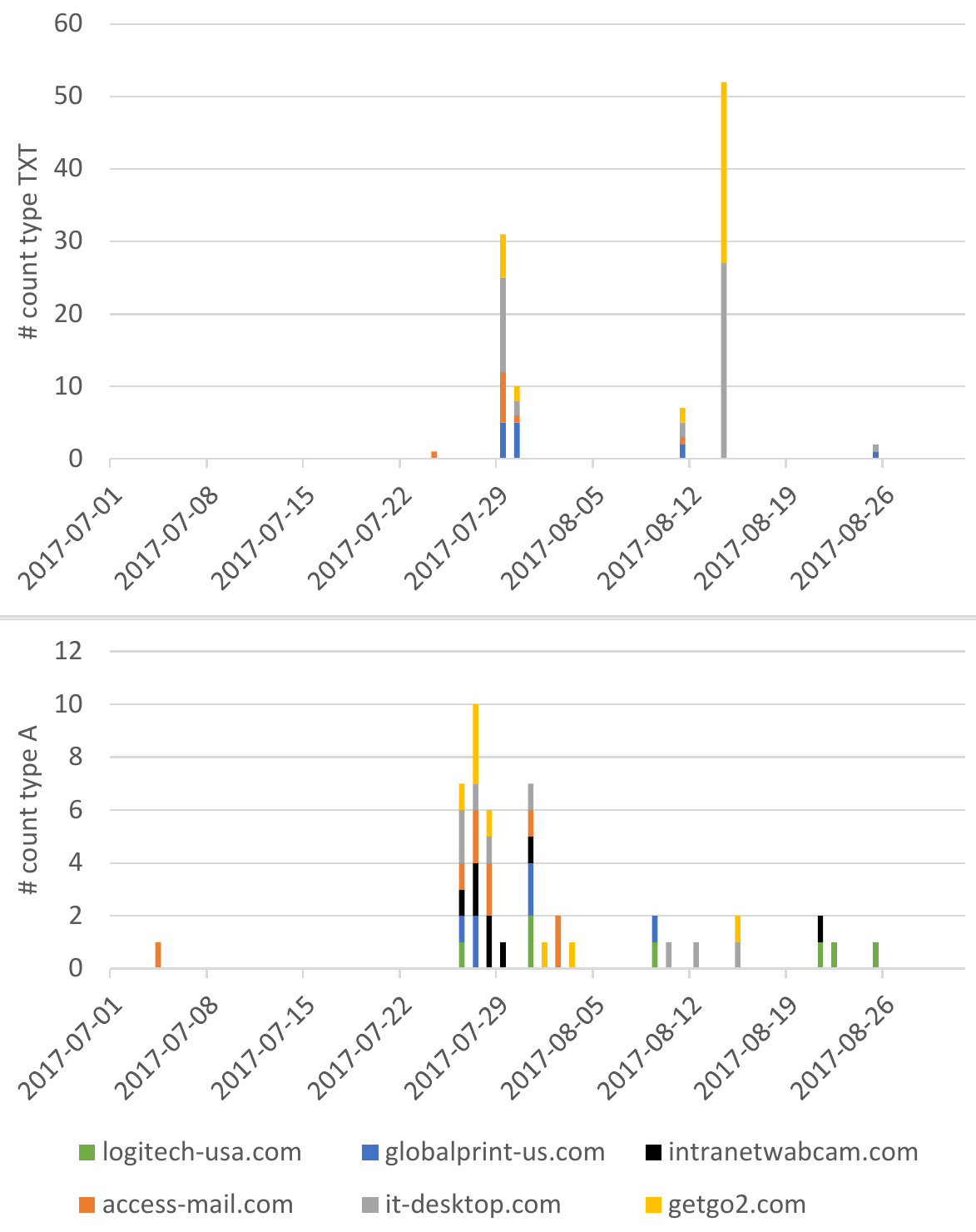}
  \caption{Wekby activity over time for rrtype TXT and rrtype A}
  \label{fig:wekby_distribution}
\end{figure}

Wekby is a second APT group which used a DNS C2 communication channel in mid-2016~\cite{wekby}. 
It is remarkable that since that time no further evidence exists on the use of this communication channel.
For this reason, one might believe that the APT group---or to be precise, the infrastructure used for the particular campaign---is not active anymore.
Nevertheless, we observed with our global view of DNS requests with new FQDNs that the DNS C2 infrastructure of the Wekby group has been used two times in our measurement period.
This, in turn, means that the covert channel is still active.
We discovered domains belonging to Wekby with two different resource record types (A and TXT).
Figure~\ref{fig:wekby_distribution} shows the activities of the known Wekby second-level domains in our data as a stacked bar chart.
The infrastructure was used two times, once between 17/07/26 and 17/07/31 and the second time between 17/08/10 and 17/08/25. 

\section{Threats To Validity}
\label{sec:limitations}

In the following, we discuss several threats to validity and limitations of our work. 
The first restriction of our work is the focus on Farsight only as provider of pDNS data.
Thus, our global view is basically Farsight's view on the DNS ecosystem.
Nevertheless, as far as we know, Farsight offers the most comprehensive and complete view of DNS usage through pDNS data.
They further advertise to provide the largest real-time actionable threat intelligence on Internet changes~\cite{farsight_noh}, and previous work also used Farsight SIE for global views~\cite{gao2013empirical}.

A further limitation is that we could not directly compare our filter approach with other DNS tunnel detection approaches. However, a comparison without further adaptations makes little sense, because we use a unique vantage point, while existing work typically evaluated their approach on internal networks. In our study, we examined DNS tunneling from a global perspective (Farsight's view) and not just locally so that a comparison is not feasible. Additionally, to our knowledge neither implementations nor data sets of existing works are available for a direct comparison.

Furthermore, the approach to identify DNS tunneling through the analysis and filtering of pDNS data has limitations. 
Unfortunately, it is not possible to assign the domain candidates to DNS tunnels with 100~\% certainty, as we do not have any ground truth on a global scale. 
To the best of our knowledge, however, the indicators and behavior are most likely DNS tunnel traffic.

The filter approach allows DNS tunnels that do not transmit a lot of data, to be overlooked and wrongly filtered. Our method is therefore only capable of robustly identifying larger data transmissions inside hostnames from level four and higher.
Potential attackers may exploit this fact and can, therefore, bypass our filter steps. However, data transmission with low bandwidth are more difficult because data can no longer be transmitted with bandwidth as large as possible. Otherwise, we would detect the tunnel activity.

The assignment of domain candidates to DNS tunnel implementations is limited in the sense that we cannot confirm that the candidates use the identified DNS tunnel implementation. 
However, we can say that we find very similar structures and thus have made the correct allocation with high probability. 
Of course, an attacker could also use custom implementations that bypass our heuristics.

The confirmation that domains have been maliciously exploited is based on news and blog posts and may not be complete. 
However, we tried to collect the information as systematically and thoroughly as possible.

Another constraint is that pDNS data analysis inspects machine-to-machine communication only, i.e., in turn, we do not know who is using a DNS tunnel and who is under attack. However, since we wanted to learn in a first step whether DNS tunnels are used for evil purposes at all in the wild, this is not within the scope of our work. In future work, one may try to encode the queries identified as DNS tunnels and thus determine the content to draw further conclusions about the individual use of the particular tunnel. 

Finally, manual work must always be included to validate the filtered domain candidates. However, this is fine, since the number of candidates should be manageable for manual analysis. 
If it is not the case, it is possible to miss potential DNS tunnel domains.

\section{Ethical Considerations}
Since we have only used pDNS data for our analyses, we have not stored or analyzed any personal data. This is only the machine-to-machine communication of DNS servers. Furthermore, we have made no effort to analyze the transmitted data to identify potential senders or receivers or to learn what information was transmitted.

\section{Related Work}
\label{sec:related_work}

Work related to ours can be divided into three categories. First, papers dealing with the detection of malicious domain names, second measurement studies in the context of DNS, and third articles in the field of DNS tunnels.

\subsubsection*{Detection of malicious domain usage}
Past publications already suggested systems to identify malicious domains based on DNS information.
Antonakakis et al. introduced Notos~\cite{antonakakis2010building}, which analyzes pDNS data to detect a  malicious domain based on statistical features like the number of IP addresses previously assigned to the domain or the number of malware samples which reached out to the domain.
Bilge et al. proposed another system called Exposure~\cite{bilge2011exposure}, which uses a similar approach but needs less training time and classifies domains correctly, which were misclassified by Notos.
These DNS reputation systems focus on characteristics of the domain itself or its usage.
However, using DNS tunnels leads to different patterns which these systems cannot detect.
Additionally, usage of a DNS tunnel is independent of the maliciousness of the underlying domain so that it is not helpful.

Furthermore, Antonakakis et al. presented Kopis~\cite{antonakakis2011detecting}, a system to detect malicious domains at a higher level of the DNS hierarchy than Notos and Exposure.
Thereby, they achieve a global view and earlier detection of malicious domains.
Liu et al. used pDNS data analysis to detect the usage of subdomains for malicious purposes~\cite{Liu:2017:DLO:3133956.3134049}.
In this technique, referred to as shadow domains, malicious actors gain access to legitimate domains, e.g., via phishing.
Afterward, they register additional subdomains, which benefit from the reputation of the original domain when used.
Compared to our work, the detection mechanism's approach is similar.
However, our detection mechanism takes individual characteristics of DNS tunnels into account, e.g., the type of the resource record or the length of the rdata field.

\subsubsection*{Measurement studies}
Many measurements have already investigated various aspects of DNS.
Examples of recent work are a study on interceptions~\cite{217551}, a study on censorship~\cite{pearce2017global}, a study on dependencies~\cite{Dell'Amico:2017:LMM:3134600.3134637}, and about measurement challenges~\cite{7460220}.
However, we are unfamiliar with any work that takes a more detailed view on newly observed hostnames.

\subsubsection*{DNS tunnels}
Several papers previously dealt with the detection of DNS tunnels~\cite{qi2013bigram,homem2017harnessing, aiello2015dns,farnham2013detecting,dusi2009tunnel,sheridan2015detection,DBLP:journals/corr/abs-1004-4358,nuojua2017dns,satam2015anomaly,born2010ngviz,ellens2013flow,cejka2014stream,karasaridis2006,aiello2013basic}.
Homem and Papapetrou presented a machine learning approach to discover protocols being tunneled within the DNS~\cite{homem2017harnessing}.
Qi et al. described a bigram based approach to detect DNS tunnels among regular DNS traffic~\cite{qi2013bigram}.
Aiello et al. presented a DNS tunnel detection technique based on statistical fingerprints of DNS packet sizes as well as the time-interval in between~\cite{aiello2015dns}.
Accordingly, various works exist; however, these works always require an internal network view.
No work uses pDNS data to analyze the usage of DNS tunnels in the wild.

In particular, the work of Paxson et al.~\cite{paxson} is similar to our work with regard to the identification of DNS tunnels. 
The authors introduced a technique to identify DNS tunnels using a configurable threshold of the amount of information within an FQDN.
For that, they presented a procedure to measure the information content of DNS query streams.
They evaluated and determined this method empirically and were able to detect 59 confirmed tunnels (2 from an enterprise network with individual clients and 57 from aggregated clients).
In addition to enterprise networks, they also used data from Farsight (SIE) for aggregated data.
The first difference is that we first present a study on newly observed hostnames.
And even with the filter steps, we utilize simpler attributes and filter functions to detect DNS tunnels in our data. We do not measure the information content in query streams, but only use attributes such as the number of subdomains or the resource record type.
Another difference is that with our measurement study, we examined the use of only new FQDN and focused on the global utilization. In addition, we analyzed the malicious use of two confirmed DNS tunnels in separate case studies.

Other publications already analyzed malware using DNS tunnels for data exfiltration or C2 communication~\cite{dietrich2011botnets, binsalleeh2014characterization, xu2013dns, kara2014detection}.
However, no paper deals with the worldwide use of DNS tunnels for malicious purposes.
\section{Future Work}
\label{sec:futurework}

We had only access to two months of data, while our method could be applied to larger data sets.
Therefore, for future work, it is worth enlarging the analysis period by buying access to the data feed to better understand the temporal evolution of the use of newly observed hostnames.
Additionally, an analysis of another second pDNS source would be interesting to see if the results change.

Since our results show that DNS tunnels are responsible for a significant proportion of newly observed hostnames, further analyses in the field of DNS tunnels might also be interesting.
In particular, a comparison of different approaches to the identification of DNS tunnels could be performed.
It might also be of interest to investigate the transmitted data to determine not only the actual usage but also the reason for DNS tunneling. 
Last, the analysis of malware samples using DNS tunnels may also be a promising option in the future.

\section{Conclusion} 
\label{sec:conclusion}

In this paper, we presented new insights into the usage of newly observed hostnames in the DNS via an empirical measurement study.
We showed that a small amount of second-level domain names are responsible for a significant fraction of the total amount of newly observed new hostnames every day.
In particular, Google's AMP project, a DNS tunnel provider, and Spotify are responsible for about half of all requests with new hostnames on the Internet.

Furthermore, we demonstrated that it is possible to identify DNS tunnels by analyzing passive DNS data feeds. 
We found that the use of DNS tunneling is widespread and represents a large proportion of type NULL and TXT requests.
During the filtering, according to type NULL, we were even able to assign the remaining domain candidates to DNS tunnel tools.
The most used tool is iodine, and a large part of the total DNS tunnel traffic belongs to the DNS tunnel provider \textit{yourfreedom.com} from Germany.
With these results, we could show that DNS tunneling is used in the wild and accounts for a considerable fraction of the total number of DNS new hostname queries.
According to our findings, DNS requests in particular of type NULL should be blocked as they are almost entirely tunneling traffic.

\section*{Acknowledgment}

This work was supported by the Office of the Director of National Intelligence (ODNI) and
the Intelligence Advanced Research Projects Activity (IARPA) via the Air Force Research
Laboratory (AFRL) contract number FA8750-16-C-0112. The U.S. Government is authorized to
reproduce and distribute reprints for Governmental purposes notwithstanding any copyright
annotation thereon. Disclaimer: The views and conclusions contained herein are those of the
authors and should not be interpreted as necessarily representing the official policies or
endorsements, either expressed or implied, of ODNI, IARPA, AFRL, or the U.S. Government.

Additionally, this work was partially supported by the Germany Federal Ministry of Education and Research (BMBF grant 16KIS0395 ``secUnity'').

% references section
\bibliographystyle{IEEEtran}
\bibliography{bibliography}

\end{document}